\documentclass[a4paper,twocolumn,11pt,accepted=2026-11-22]{quantumarticle}
\pdfoutput=1
\usepackage[utf8]{inputenc}
\usepackage[english]{babel}
\usepackage[T1]{fontenc}
\usepackage{amsmath}
\usepackage{hyperref}
\usepackage{amssymb}
\usepackage{graphicx} %
\usepackage{xcolor}

\usepackage{tikz}
\usepackage{amsthm}
\usepackage{lipsum}
\usepackage{braket}
\usepackage{stmaryrd}
\usepackage{booktabs} 

\newtheorem{definition}{Definition}

\usepackage{xspace}

\usepackage{newfloat}
\DeclareFloatingEnvironment[
    fileext=loa,
    listname={List of Algorithms},
    name=Algorithm,
    placement=tbhp,
]{algorithm}
\usepackage{algorithmicx}
\usepackage{algpseudocode} %

\usepackage{booktabs}
\usepackage{subcaption}
\usepackage{threeparttable} %

\newcommand{\ourtool}{ScaLER\xspace}
\newcommand{\stim}{Stim\xspace}

\def\ket#1{|#1\rangle}

\def\wsweet{w_{\text{sweet}}}
\def\werr{w_{\text{err}}}
\def\wsat{w_{\text{sat}}}
\def\maxshots{S_{\max}}
\def\maxerrors{N_w^{LE}}

\begin{document}

\title{Scalable testing of quantum error correction}

\author{John Zhuoyang Ye}
\author{Jens Palsberg}
\affiliation{University of California, Los Angeles}
\orcid{0000-0003-4747-365X}
\maketitle

\begin{abstract}
The standard method for benchmarking quantum error correction is Monte Carlo sampling. However, direct estimation of rare logical failures through unconditional Monte Carlo sampling can be prohibitively expensive, even with efficient simulators such as \stim. In this paper, we present a scalable approach that combines stratified fault injection with extrapolation.
Our insight is that some of the fault space can be sampled efficiently, after which extrapolation is sufficient to complete the testing task.
As a result, our tool scales to distance 17 for a physical error rate of 0.0005 with a two-hour time budget on a desktop, while Monte Carlo sampling using \stim observed 0 logical errors at distance 13 with the same time budget.  At distance 17, ScaLER estimated a logical error rate of $1.51 \times 10^{-11}$ with high confidence.
\end{abstract}

\tableofcontents

\section{Introduction}
\label{sec:introduction}

\paragraph{Background.}

Quantum computing has made rapid progress this decade, including a breakthrough in 2024 when researchers demonstrated that {\em quantum error correction (QEC)}
\cite{reichardt2024logical,Bravyi_2024,rodriguez2024experimentaldemonstrationlogicalmagic}
can partially suppress quantum-hardware errors \cite{GoogleBreakEven}.
This breakthrough paves the way for fault-tolerant quantum computing, which is expected to become a reality in the next decade and begin to have an impact on science and business \cite{shor1999polynomial,HHL,grover}.

Experimental papers on QEC often express the quality of their approaches by stating the {\em logical error rate}, which is the probability that a program execution gives a wrong result \cite{Quantinuum2024}.
Typically, such papers estimate the logical error rate through {\em testing} of the quantum circuit that implements the QEC algorithm \cite{geher2024error,anbang,Domokos2024characterizationof}.
Most of these circuits are too large for current quantum computers, so instead the authors simulate them on classical computers, using a model of quantum-hardware errors \cite{Tomita_2014}.
Thus, the benchmarking challenge boils down to a testing campaign on a classical computer with the goal of estimating the logical error rate in a repeatable way.

For high-quality QEC algorithms with low logical error rates, the testing required is massive and has become a major obstacle for researchers
\cite{gidney2021stim,BenchmarkPlanar,geher2024error,FaultTolerantPreparation,gidney2023yokedsurfacecodes}.
Furthermore, the lower the logical error rate, the more testing is needed to estimate what it is
\cite{Gidney2023pairmeasurement,Heu_en_2024, beverland2025failfasttechniquesprobe}.
The goal of this paper is to greatly reduce the testing effort for high-quality QEC algorithms.

\paragraph{State of the Art.}

Unconditional Monte Carlo sampling using Gidney's popular simulator \stim \cite{gidney2021stim} applies
randomized fault injection testing \cite{FITT} to benchmark an implementation of quantum error correction.
Monte Carlo sampling using \stim repeatedly injects random faults into a circuit and checks for logical errors. However, for high-quality QEC algorithms, logical errors are rare events, requiring unconditional Monte Carlo sampling to run a massive number of tests to obtain a statistically reliable estimate of the logical error rate.
For example, for a surface code with distance 13 and for a physical error rate of $0.0005$, Monte Carlo sampling using \stim found no logical errors at all after running for two hours on a desktop.
This motivated us to design a more scalable approach.

\paragraph{Our Results.}
We present \ourtool (Scalable Logical Error Rate Testing), a scalable approach to benchmarking implementations of quantum error correction.
Our insight is that some of the fault space can be sampled efficiently, after which extrapolation is sufficient to complete the testing task.
Our approach complements unconditional Monte Carlo sampling using \stim by estimating rare logical failures at large code distances. Our tool produces repeatable results for all distances we have tried and, for small distances, consistently matches Monte Carlo sampling using \stim in order of magnitude, although individual estimates can differ by as much as 60\%.

For this time budget, \ourtool estimated a logical error rate of $1.51 \times 10^{-11}$. In contrast, for the same time budget, the lowest logical error rate that Monte Carlo sampling using \stim estimated with high confidence was $5.95 \times 10^{-6}$ for a surface code with distance 7.

\paragraph{The Rest of the Paper.}
In Section~\ref{sec:examples-of-how-stim-works}, we give examples of how Monte Carlo sampling using \stim works,
in Section~\ref{sec:modeling-qec-test-data}, we discuss how to model QEC-test data, and
in Sections~\ref{sec:separating-low-and-high-weights}--\ref{sec:our-algorithm}, we present our algorithm.
In Section~\ref{sec:implementation}, we present our implementation, 
in Section~\ref{sec:evaluation}, we detail the evaluation of our approach, and 
in Section~\ref{sec:threats-to-validity}, we discuss threats to validity.  
Finally, 
in Sections~\ref{sec:related-work}--\ref{sec:future-work} we discuss related and future work, and 
in Section~\ref{sec:conclusion}, we conclude.

Our software is publicly available under an open-source license at this link:

\href{https://github.com/yezhuoyang/ScaLERQEC}{https://github.com/yezhuoyang/ScaLERQEC}.

\section{Examples of Monte Carlo Sampling Using \stim}
\label{sec:examples-of-how-stim-works}

In this section, we recall the relationship between distance and logical error rate, and we discuss how unconditional Monte Carlo sampling using \stim works and why it becomes sample-inefficient when logical failures are rare.  Our examples are a repetition code with distance 3 and a surface code with distance 7.

\subsection{Distance and Logical Error Rate}

For QEC implementations, a higher distance means a lower logical error rate.
In other words, larger scale means better quality.
Medium-scale distances are 5--13, while large-scale distances range from 15 and up, with 27 as a particularly well-studied point 
\cite{Gidney2021howtofactorbit}.
A QEC algorithm with distance $d$ can correct $(d-1)/2$ errors, so higher is better, but what if more than $(d-1)/2$ errors occur?  
In some cases, a QEC algorithm successfully handles such cases, while in others, the results are logical errors.  Researchers summarize a testing campaign by estimating the logical error rate as the number of logical errors divided by the number of tests.

\subsection{Repetition Code}

\begin{figure}[t]
    \centering
    \includegraphics[width=0.45\textwidth]{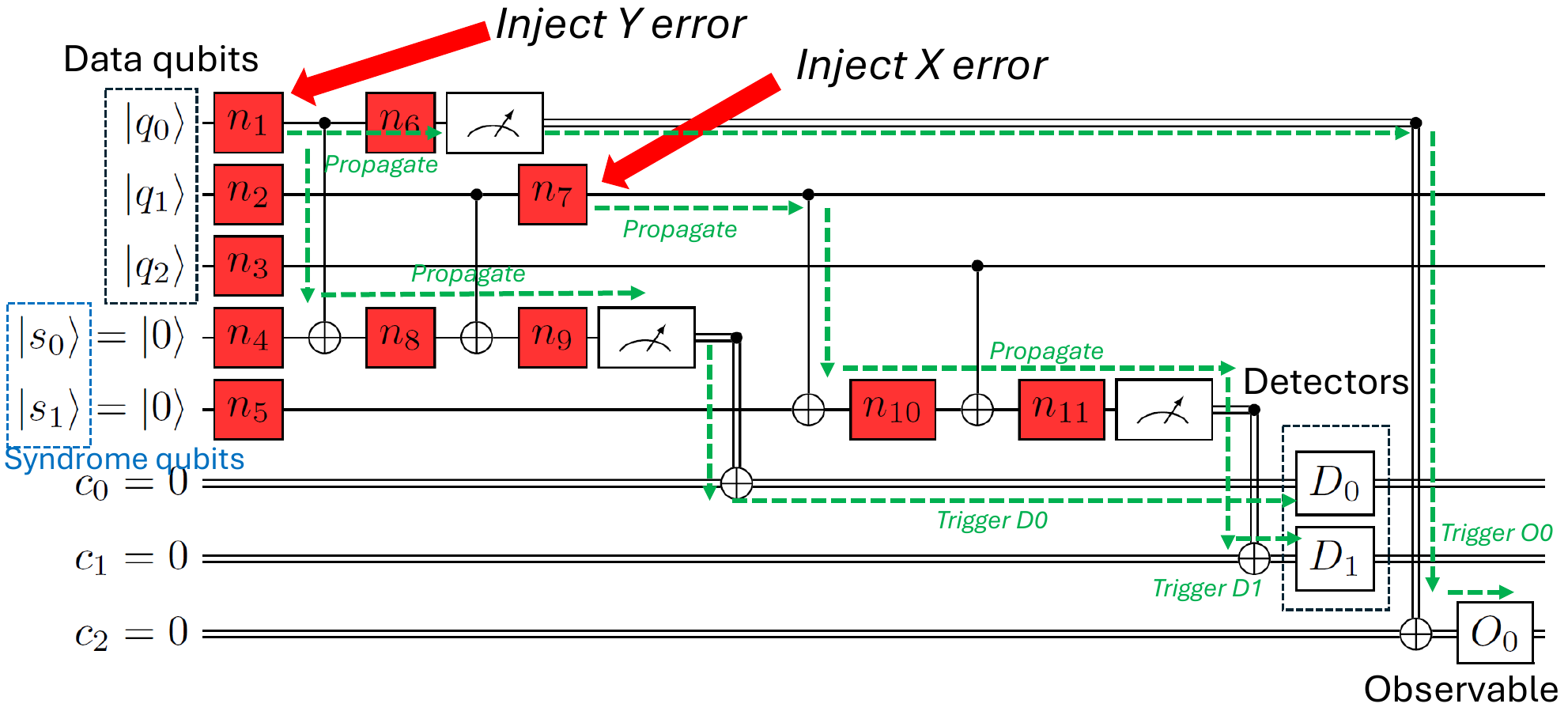}
    \caption{Diagram for injecting SID error model to a $[[3,1,3]]$ repetition code circuit which protect logical $\ket{0}$ state. There are $11$ error locations indexed by $n_i$, $i \in [1,11]$. We inject fault $n_7=X$, which will trigger detector $D_1=1$ after propagation. We also inject fault $n_1=Y$, which trigger detector $D_0=1$ and also flip observable result $O_0=1$.}
    \label{fig:circuitProp}
\end{figure}

Let us see how Monte Carlo sampling using \stim estimates the logical error rate for the circuit in Figure~\ref{fig:circuitProp}, which implements a repetition code with distance 3.
Specifically, it encodes a source-level qubit as three data qubits $\ket{q_0},\ket{q_1},\ket{q_2}$, uses two detectors $D_0,D_1 \in \{0,1\}$ for mid-circuit error detection, and lets $O_0 \in \{0,1\}$ denote the result of measuring the encoded qubit.
\stim models quantum-hardware errors by repeatedly injecting faults into the circuit and then testing whether the circuit still works correctly.   
Conceptually, \stim does that by:
\begin{enumerate}
    \item picking a random list of circuit locations,
    \item injecting an X-gate, Y-gate, or Z-gate into each of those locations, and
    \item testing whether the modified circuit works correctly, with the help of a decoder.
\end{enumerate}
We will refer to the modified circuit as a {\em sample}.  If the list of locations has length $w$, then we say that the sample has weight $w$.
Our example circuit in Figure~\ref{fig:circuitProp} has 11 locations in which \stim can inject faults. Here, we restrict the noise model to Pauli-$X$ errors only, ignoring $Y$ and $Z$ errors.
Table~\ref{tab:Example1} summarizes a testing campaign with 11 samples, all with weight 1.
\stim found that 2 of the 11 samples give a logical error, while the rest work correctly.
From this testing campaign, our estimate of the logical error rate $P_L$ is: 
\begin{eqnarray}
\label{eq:estimate-of-logical-error-rate-repetition-code-3}
P_L 
\;\approx\;
\frac{\mbox{number of logical errors}}{\mbox{number of samples}}
\;=\;
\frac{2}{11}
\end{eqnarray}

\begin{table}[t!]
\centering
\resizebox{\columnwidth}{!}{%
\begin{tabular}{|l|l|l|l|l|l|l|l|}
   \hline 
 Index & \textbf{Error} & $D_0$ & $D_1$ & $O_0$ & Decoder output& Predicted $O_0$ & Logical error?  \\
  \hline
 1 & $n_1=X$ & 1 & 0 & $1$ & $n_4=X$ &  $0{(\color{red} \neq 1)}$ &  Yes\\
  \hline
2  & $n_2=X$ & 1 & 1 & $0$ & $n_2=X$  & $0$ &  No \\
  \hline
3  & $n_3=X$ & 0 & 1 & $0$ & $n_3=X$  & $0$  & No \\
  \hline
4  & $n_4=X$ & 1 & 0 & $0$ &  $n_4=X$   & $0$ & No  \\
  \hline
5  & $n_5=X$ & 0 & 1 & $0$ &  $n_3=X$    & $0$  & No  \\
  \hline
6  & $n_6=X$ & 0 & 0 & $1$ &  No Error   & $0{(\color{red} \neq 1)}$ &  Yes  \\
  \hline
7  & $n_7=X$ & 0 & 1 & $0$ &  $n_3=X$     & $0$ &  No  \\
  \hline
8  & $n_8=X$ & 1  & 0  &  0 &   $n_4=X$     &  $0$   &  No  \\
  \hline
9  & $n_9=X$ & 1  & 0  & 0  &  $n_4=X$    & $0$  &  No  \\
  \hline
10  & $n_{10}=X$ & 0  & 1 & 0 & $n_3=X$      & 0 & No   \\
  \hline
11  & $n_{11}=X$ & 0  & 1 & 0 & $n_3=X$      & 0  & No   \\
  \hline
\end{tabular}
}
\caption{All $11$ cases with one Pauli X error when the logical state is $0$. The outcomes of the detector/observable are calculated by tracking the propagation of the injected error.}
\label{tab:Example1}
\end{table}

\subsection{Surface Code}
\label{sec:SurfaceCode}

\begin{figure}[t]
    \centering
    \includegraphics[width=0.45\textwidth]{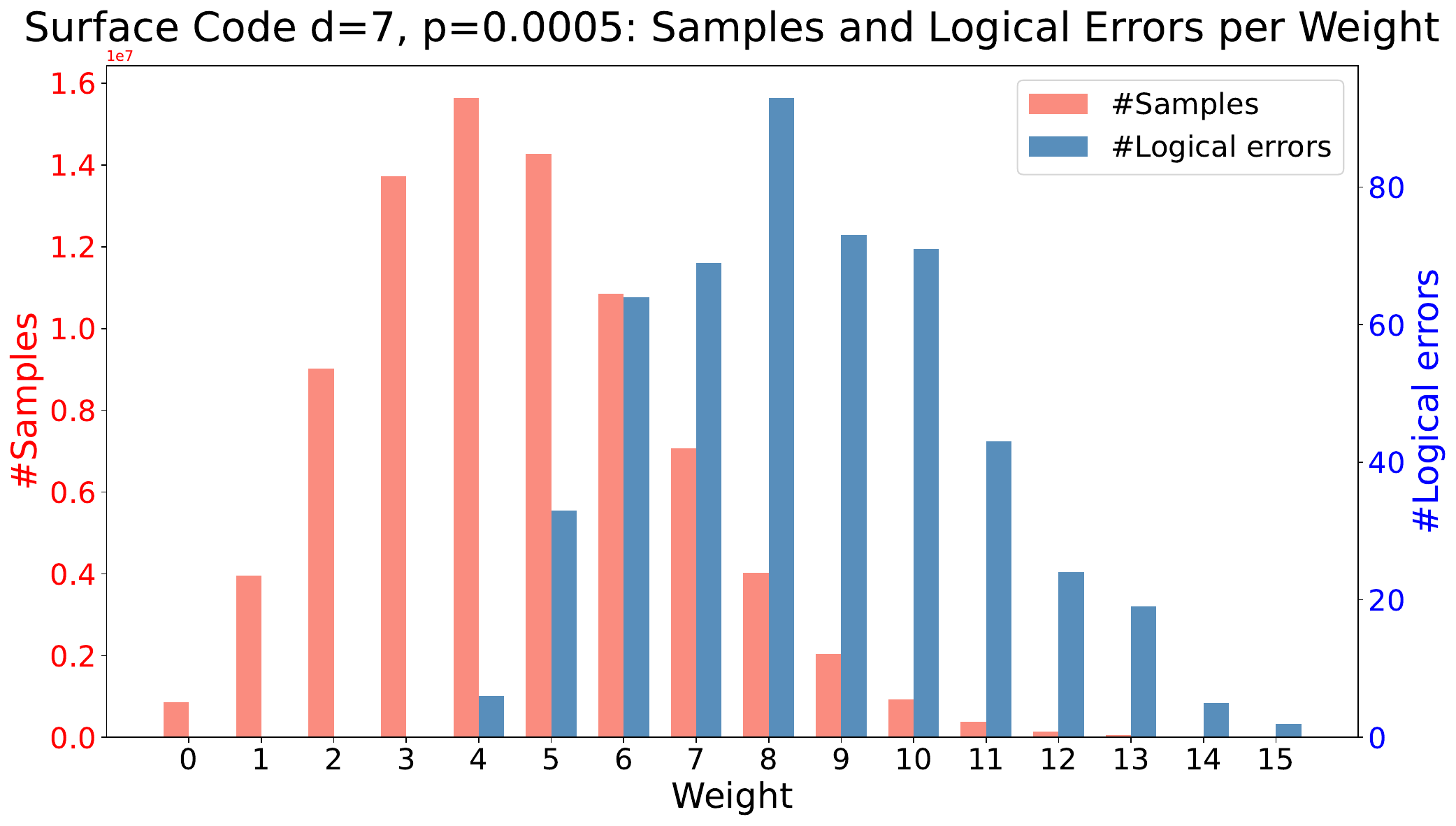}
    \caption{We run Monte Carlo sampling using \stim on Surface code with distance $7$ and physical error rate $5\times 10^{-4}$ and plot the number of samples ({\color{red} red bars}) and the number of logical errors ({\color{blue} blue bars}) for each weight. Below $w=3$, the circuit is fault-tolerant and Monte Carlo sampling using \stim cannot sample any logical errors. Beverland et al. and Carolyn Mayer et al. report similar observations in their recent papers \cite{mayer2025rareeventsimulationquantum,mayer2025rareeventsimulationquantum}. }
    \label{fig:bar-diagram-number-of-samples-for-each weight}
\end{figure}

Let us see why unconditional Monte Carlo sampling using \stim becomes sample-inefficient when logical failures are rare. We added lightweight logging hooks to \stim to record the Pauli-weight for every sample. Figure~\ref{fig:bar-diagram-number-of-samples-for-each weight} summarizes the testing campaign of Monte Carlo sampling using \stim for a surface code with distance 7, showing, for each weight, the number of samples and the number of logical errors.
The low-weight samples dominate, which is due to the error model that we use. Our model is a variant of a popular error model called SD6 \cite{Gidney2021faulttolerant,geher2024error}.  We call our model SID, which is short for the \emph{uniform single-qubit circuit-level independent depolarizing noise model}. This model has a single parameter $p$, known as the {\em physical error rate}, which is the probability that a gate, a measurement, or an idling step works incorrectly due to depolarization noise. 
The model assumes that such errors are independent of each other and that each error can be modeled by injecting an X-gate, Y-gate, or Z-gate. This implies that the model excludes correlated Pauli errors \cite{aharonov2006fault}.

\begin{quotation}
\noindent
{\em We use the SID model with $p = 0.0005$, unless noted explicitly.} 
\end{quotation} 
This models some of the best quantum-hardware today, taking into account circuit-level behavior such as measurement/reset error and hook errors
\cite{reichardt2024logical,Quantinuum2024,2023Surface}. In addition, $p=0.0005$ is better than the threshold required for quantum error correction to be successful. In Section~\ref{sec:evaluation}, we experiment with $p$ down to $p=0.0001$ to stress test our method. 

In the SID error model, \stim uses $p$ to determine a weighted binomial distribution (Figure~\ref{fig:bar-diagram-number-of-samples-for-each weight}) centered on $p$ multiplied by the number of locations in the circuit.
The circuit for the surface code with distance 7 has 9,121 locations, so \stim centers the weight distribution around $0.0005 \times \mbox{9,121} \approx 4$.
In total, Monte Carlo sampling using \stim tried 83,000,000 samples and found 503 logical errors, from which we can estimate the logical error rate:
\begin{eqnarray*}
P_L 
&\approx&
\frac{\mbox{number of logical errors}}{\mbox{number of samples}} \label{eq:estimate-of-logical-error-rate-surface-code-7} \\[2pt]
&=&
\frac{503}{\mbox{83,000,000}} %
\;\;=\;\;
6.06 \times 10^{-6}
\end{eqnarray*}

Figure~\ref{fig:bar-diagram-number-of-samples-for-each weight} shows that the proportion of logical errors sampled is low at low weights, but higher at higher weights.
This crystallizes the reason why unconditional Monte Carlo sampling using \stim becomes sample-inefficient when logical failures are rare: it tests mostly low-weight samples, which have few logical errors.
Indeed, for our example, Monte Carlo sampling using \stim spent $8.22$ minutes on low-weight samples (weights 0--8) but only $0.37$ minutes on high-weight samples (weights 9--15).
When the distance increases, the threshold theorem \cite{gottesman1998theory,shor1996fault} says that the logical error rate decreases exponentially.  Our experiments with Monte Carlo sampling using \stim confirm this: when we increased the distance of the surface code from 7 to 13, the number of logical errors decreased from 503 to 0.
If we were to use this outcome to estimate the logical error rate, we would get 0, which is wrong.

In this paper, we do the opposite of Monte Carlo sampling using \stim: we test only high-weight samples and use the results to predict the results for low-weight samples.
This scales well because high-weight samples readily produce logical errors and because we have an efficient prediction method.

\section{Modeling QEC-Test Data}
\label{sec:modeling-qec-test-data}

In this section, we show how to model the data from any QEC-testing campaign.
Such a model enables us to use the results for high-weight samples to predict the results for low-weight samples.

\paragraph{Stratified Logical Error Rates.}

Our first step is to reformulate the definition in Equation~\ref{eq:estimate-of-logical-error-rate-repetition-code-3} of how to estimate the logical error rate from the QEC-test data, as in Figure~\ref{fig:bar-diagram-number-of-samples-for-each weight}.
For each weight $w$, we can calculate the logical error rate $P^w_L$ for that $w$, as shown in Equation~\ref{eq:probability-of-logical-error-at-weight-w}.
This enables us to sum up those probabilities $P^w_L$, weighted by the probability that $w$ faults occur, as shown in 
Equation~\ref{eq:stratified-estimate-of-logical-error-rate}.
\begin{eqnarray}
P^w_L &=& 
P(\mbox{logical error at weight } w) 
\nonumber \\
&\approx& 
\label{eq:probability-of-logical-error-at-weight-w}
\frac{\mbox{\# logical errors at weight $w$}}{\mbox{\# samples at weight $w$}}
\\
P_L &=&
P(\mbox{logical error}) 
\nonumber \\
&\approx&
\label{eq:stratified-estimate-of-logical-error-rate}
\sum_{w}
P^w_L \times P(w \mbox{ faults occur})
\end{eqnarray}

For a physical error rate $p$ and a circuit with $C$ locations, we have:
\begin{eqnarray}
P(w \mbox{ faults occur}) &=&
\binom{C}{w} p^w(1-p)^{C-w}
\label{eq:FinalProof}
\end{eqnarray}
Equation~\ref{eq:FinalProof} is correct for the error model we use because for a weight $w$, 
there are $3^w\binom{C}{w}$ different error configurations that each occur with probability $(\frac{p}{3})^w(1-p)^{C-w}$. 
Notice that $3^w$ and $(\frac{1}{3})^w$ cancel each other out.

\paragraph{The Logical Error Rates Form an S-Curve.}

\begin{figure}[t]%
    \centering
    \includegraphics[width=0.43\textwidth]{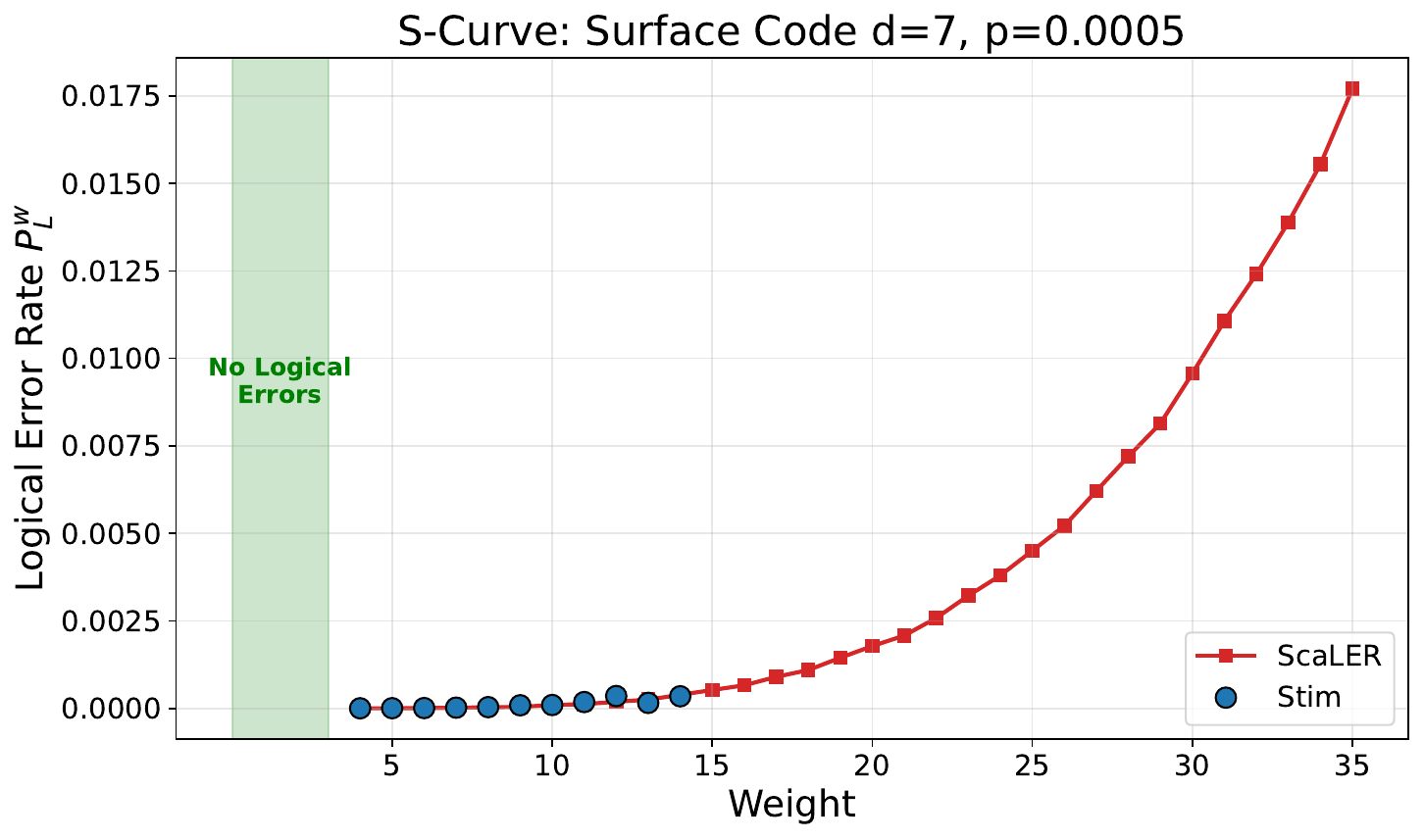}
    \caption{S-curve data (Surface code $d{=}7$, $p=5\times 10^{-4}$) combining Monte Carlo samples generated using \stim and \ourtool extensions.}
    \label{fig:ScurveData-separate}
  \end{figure}

For the data in Figure~\ref{fig:bar-diagram-number-of-samples-for-each weight},
we can plot $P^w_L$ as a function of $w$, which produces the blue data points in Figure~\ref{fig:ScurveData-separate}.
In the left part of Figure~\ref{fig:ScurveData-separate}, for low weights, we have the fault-tolerant zone where the logical error rate is 0. This zone goes from weight 0 to $(d-1)/2 = 3$. From weights 4 to 14, we have \textcolor{blue}{blue} $P^w_L$ data points based on Figure~\ref{fig:bar-diagram-number-of-samples-for-each weight} and Equation~\ref{eq:probability-of-logical-error-at-weight-w}, and from weights 15 to 35 we have \textcolor{red}{red} $P^w_L$ data points that we calculated after using our own tool \ourtool to perform random tests for these weights.  

\begin{figure*}[t]
    \centering
    \includegraphics[width=\textwidth]{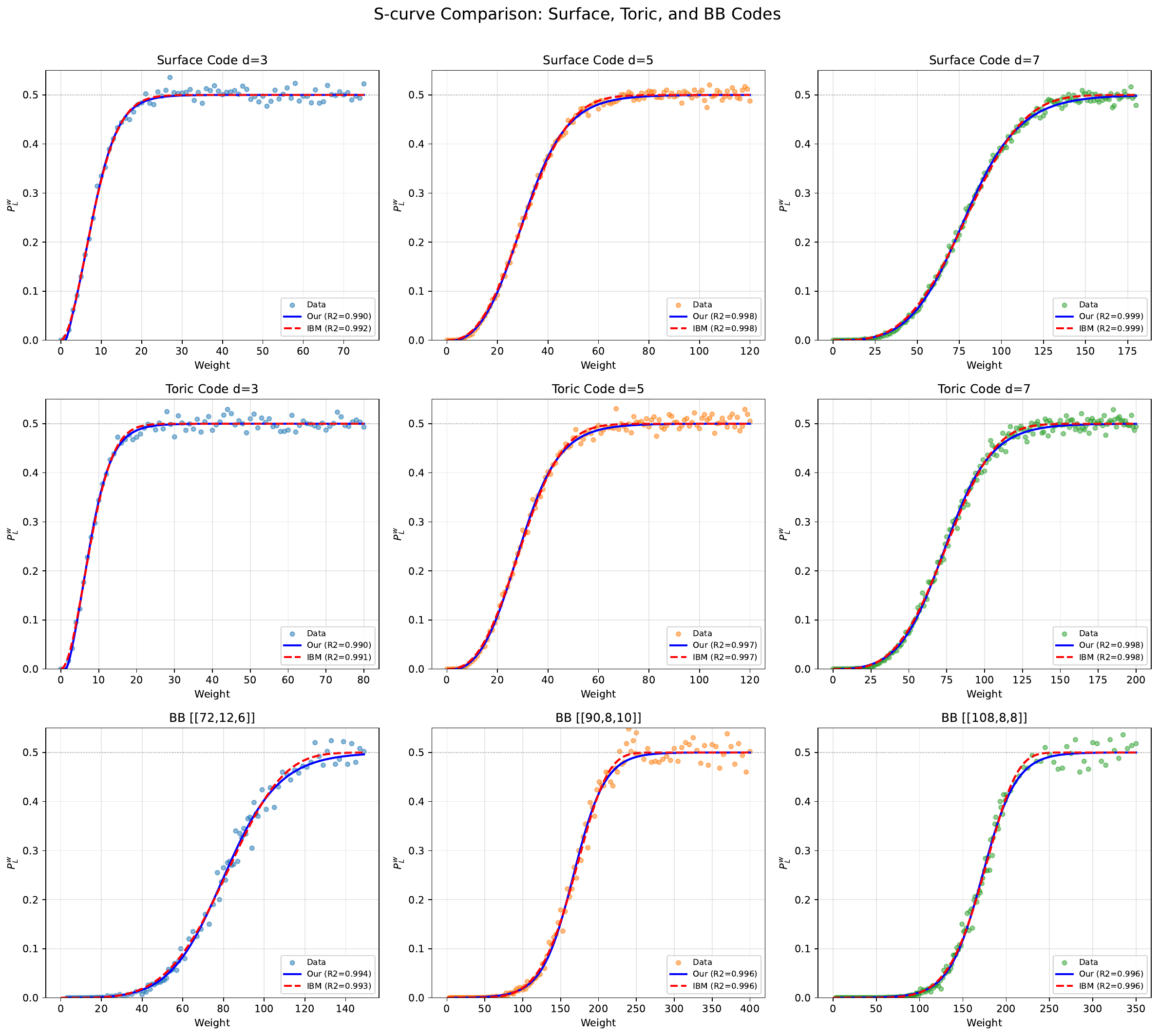}
    \caption{We fit the data collected by small scale QEC circuit (d=3,5,7) for Surface code, Toric code and Bivariate Bicycle code by both S-curve model under single qubit depolarization noise model and . Both models fits perfectly well with the observed data for all three codes. We calculate the $R^2$ score to evaluate the curve fitting. $R^2>0.99$ for all cases(Table \ref{tab:Rsquare}).}
    \label{fig:allcodes}
\end{figure*}

In Figure~\ref{fig:ScurveData-separate},
the \textcolor{red}{red} data points show a steep rise, which can make us wonder what happens for weights higher than 35.
We answer this by continuing the random testing for weights greater than 35, see Figure~\ref{fig:allcodes}.  The data forms an S-curve that flattens out at 0.5, which is where the physical errors saturate the circuit and the logical errors happen with probability 0.5, like a coin flip.
We tried this experiment with also a toric code and a BB code \cite{lee2025color}, and we tried for different distances. The results are shown in Figure~\ref{fig:allcodes}.  
In every case, we see a curve shaped as the letter S, which leads us to call them S-curves.  
We tried other cases beyond those shown in Figure~\ref{fig:allcodes} and in every case we saw an S-curve.

We want to model the S-curves that emerge from the QEC-test data.  For our purposes, the best kind of model is a continuous function.  We will use $\mathbb{R}^+$ to denote the set of nonnegative real numbers.

\begin{definition}[Modeling S-Curves]
\label{def:modeling-s-curves}
  We have the following requirements for a function $f$ that models the S-curve in QEC-test data: 
\begin{itemize}
\item $f: \mathbb{R}^+ \rightarrow \mathbb{R}^+$ and is twice differentiable (except at 0).
\item $f(0) = 0$.
\item $f$ is increasing.
\item $\lim_{w \rightarrow \infty} f(w) = 0.5$.
\item $f''$ changes sign from $+$ to $-$ at an inflection point.
\end{itemize}
\end{definition}

Previous work in many fields has defined functions that satisfy the requirements in Definition~\ref{def:modeling-s-curves}, particularly for purposes such as neural-network training \cite{han1995influence,cybenko1989approximation}, natural-phenomena modeling \cite{yin2003flexible,hassell1977sigmoid,klimstra2008sigmoid}, and prediction of software-failure \cite{khoshgoftaar1996using,iqbal2022modelling}.
Now we give a couple of specific definitions of $f$ that satisfy those requirements, and then we show that they fit the QEC-test data well. 
We begin with the first published S-curve model of QEC-test data, which appeared in a seminal paper by researchers from IBM \cite{beverland2025failfasttechniquesprobe}.

\begin{definition}[IBM's S-Curve Model]
\label{def:ibms-s-curve-model}
The paper by Beverland et al.~\cite{beverland2025failfasttechniquesprobe} from IBM defined the following ansatz function with three parameters to learn, called the Min-Fail Enclosure Model.  
The function takes as input an integer $w \geq 0$:
\begin{eqnarray}
    f_t^{\text{IBM}}[\mu,\alpha, \beta](w)
= && \nonumber \\
&& \!\!\!\! \!\!\!\! \!\!\!\! \!\!\!\! \!\!\!\!
\!\!\!\! \!\!\!\! \!\!\!\! \!\!\!\! \!\!\!\!
\frac{1}{2} \left[ 1 - \exp\!\left(
-2\, \mu \left(\frac{w}{\beta}\right)^{\!\alpha}
\right) \right]
\label{eq:IBM}
\end{eqnarray}
\end{definition}

In the paper \cite{beverland2025failfasttechniquesprobe}, $\beta$ denotes the smallest error weight that triggers a logical error, called the {\em onset\/} weight. One can easily check that $f_t^{\text{IBM}}[\mu,\alpha,\beta]$ satisfies the requirements in Definition~\ref{def:modeling-s-curves}. 

Next, we introduce our own S-curve model.
In IBM's S-curve model, the onset weight ($\beta$ in Equation \ref{eq:IBM}) is a parameter to fit or to be found through a computationally intensive and unscalable search. This led us to treat $\beta$ as a parameter to fit. (We did try hard-coding $\beta=t+1$, but this fits the QEC-data poorly.) 

We assume that we know the distance $d$ of the QEC circuit under test and therefore the fault tolerance threshold is
\(t=\left\lfloor \tfrac{d-1}{2} \right\rfloor\), and \(t+1\) is the onset weight. This makes sense for our SID error model: since no correlated Pauli errors are introduced, the mapping from injected faults to Pauli weight is faithful, and Equation~\ref{eq:FinalProof} is accurate when stated in terms of Pauli weight. Thus, we are free from the concern of determining the onset weight. 

\begin{quotation}
\noindent
{\em The onset weight $t+1$ is known.} 
\end{quotation} 

In our experiments, IBM’s S-curve model fits the observed data well, but its extrapolation to low-weight subspaces near the fault-tolerant threshold exhibits low precision in our noise model. Moreover, the contribution to the logical error rate is dominated by these low-weight subspaces, so it directly results in low accuracy in the logical error rate estimation, as we show experimentally in Section~\ref{sec:DiffScurve}. This motivated us to design a more accurate model. 

The key idea in our S-curve model is to introduce a special point at $w=t$ in the denominator of the S-curve model, encoding that, in our SID noise model, the logical error rate is exactly zero for all subspaces with $w \le t$. We also carefully model the extrapolation towards the singular point $w=t$ to achieve good agreement with the data.

\begin{definition}[Our S-Curve Model]
    \label{def:our-s-curve-model}
    For a given integer $t \geq 0$, intended to be $t = \frac{d-1}{2}$, the function $f_t^{\text{ours}}$ is described by three real parameters $\mu,\alpha,\beta$, and takes as input an integer $w \geq 0$: 
    \begin{eqnarray*} 
        f_t^{\text{ours}}[\mu,\alpha,\beta](w) &=& \\
        && \!\!\!\!\!\!\!\!\!\!\!\!\!\!\!
        \left\{
        \begin{array}{ll}
        \frac{1}{2} \cdot \frac{1}{1 + e^{-\frac{w - \mu}{\alpha} + \frac{\beta}{\sqrt{(w-t)} }} } 
        &
        t < w
        \\ \\
        0 & 
        w \leq t
        \end{array}
        \right.
        \nonumber
        \label{eq:S-curve}
    \end{eqnarray*}
    \end{definition}
Our model approximates each $P^w_L$ by $f_t^{\text{ours}}[\mu,\alpha,\beta](w)$. 
Notice that we model the fault-tolerant zone by defining $f_t^{\text{ours}}[\mu,\alpha,\beta](w) = 0$ for $w \leq t$. 
One can easily check that $f_t^{\text{ours}}[\mu,\alpha,\beta]$ satisfies the requirements in Definition~\ref{def:modeling-s-curves}. 

\paragraph{Fitting an S-Curve to QEC-Test Data.}

Given the QEC-test data in Figure~\ref{fig:bar-diagram-number-of-samples-for-each weight}, we use the least-squares method to fit an S-curve to the data and determine the parameters.  We do this for both IBM's and our S-curve models, resulting in the curves shown in Figure~\ref{fig:allcodes} and the $R^2$ values shown in Table~\ref{tab:Rsquare}, which are all close to 1.
For example, for the surface code with distance 7 (discussed in Sections~\ref{sec:introduction}--\ref{sec:examples-of-how-stim-works}), the curve-fitting produces these parameters for our S-curve model: 
$\mu = 41.71$ and 
$\alpha = 19.93$ and 
$\beta = 16.03$. 
The resulting $f_t[\mu,\alpha,\beta]$ function fits the data in Figure~\ref{fig:allcodes} almost perfectly: $R^2 = 0.9989$.
The consistent high quality of the fits across distinct code families supports the generality, robustness, and prediction power of the two S-curve models. 
We can use this to our advantage, as we consider next.

\begin{table}[t]
\centering
\resizebox{\columnwidth}{!}{%
\begin{tabular}{|c|c|c|}
\hline
\textbf{Code [[n,k,d]]} & \textbf{Our Model $R^2$} & \textbf{IBM Model $R^2$} \\
\hline
Surface [[9,1,3]]  & 0.9902 & 0.9919 \\
Surface [[25,1,5]] & 0.9984 & 0.9982 \\
Surface [[49,1,7]] & 0.9989 & 0.9986 \\
Toric [[18,2,3]]   & 0.9900 & 0.9909 \\
Toric [[50,2,5]]   & 0.9965 & 0.9966 \\
Toric [[98,2,7]]   & 0.9977 & 0.9978 \\
BB code [[72,12,6]]    & 0.9939 & 0.9935 \\
BB code [[90,8,10]]    & 0.9956 & 0.9960 \\
BB code [[108,8,8]]    & 0.9962 & 0.9963 \\
\hline
\end{tabular}%
}
\caption{$R^2$ comparison between our model and the IBM model across codes calculated from data in Figure~\ref{fig:allcodes}. This result indicates that both S-curve models are well justified to be used to learn and evaluate the quantum error correction circuits.}
\label{tab:Rsquare}
\end{table}

\section{Separating Low and High Weights}
\label{sec:separating-low-and-high-weights}

\begin{figure}[t]%
\centering
\includegraphics[width=0.43\textwidth]{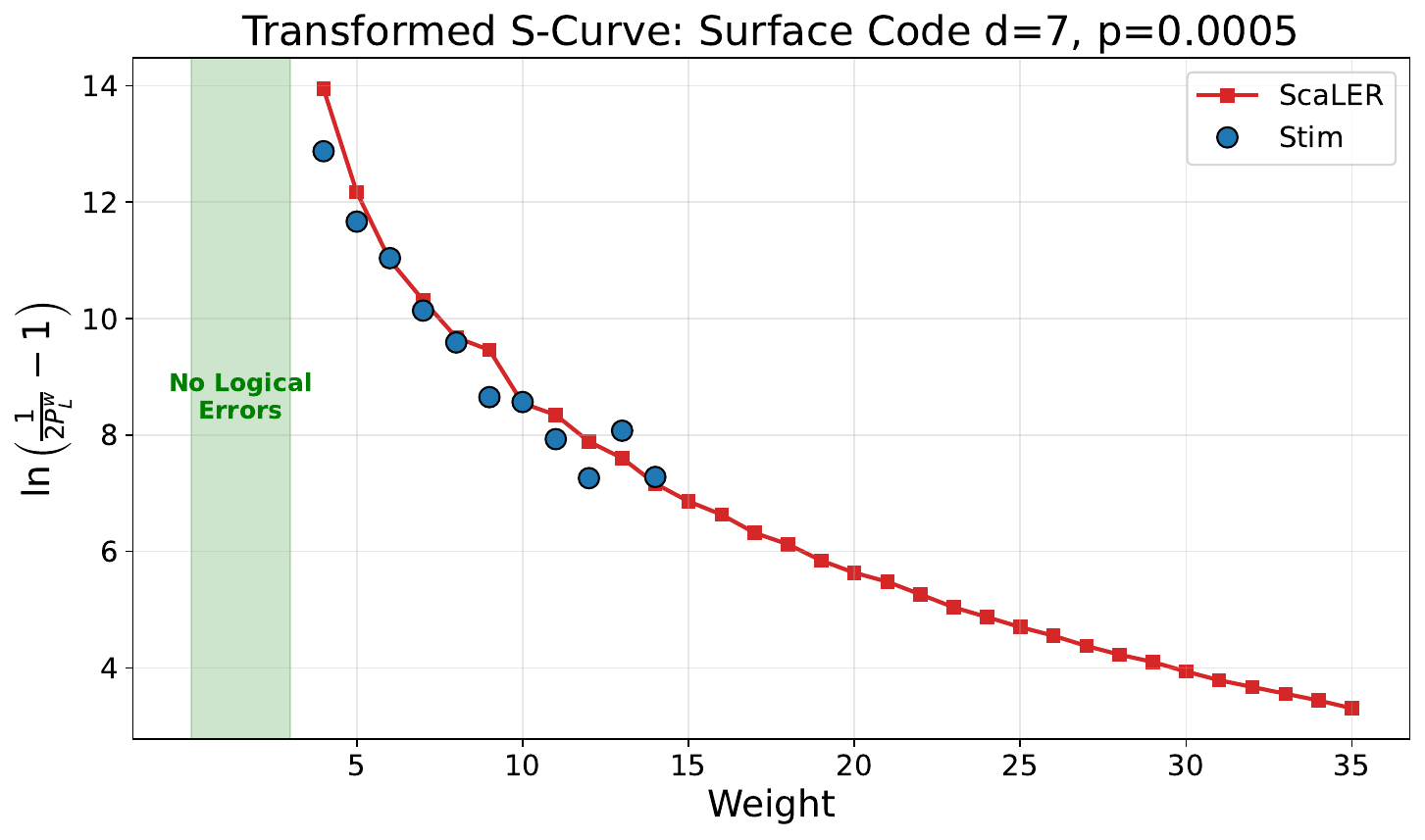}
\caption{Y-curve transformation of (c) via $\ln\!\big(\frac{1}{2P_L^w}-1\big)$ (Eq.~\ref{eq:YcurveTrans}); \textcolor{blue}{blue}: Monte Carlo sampling using \stim; \textcolor{red}{red}: \ourtool.}
\label{fig:ScurveTransformedData}
\end{figure}

In this section, we define the boundary between the low weights and the high weights, and define an upper bound on the high weights.  We call those weights $\wsweet$ (the sweet spot) and $\wsat$ (the saturation point):
\begin{eqnarray*}
1 \leq \mbox{low weights} <\wsweet \leq \mbox{high weights} \leq\wsat
\end{eqnarray*}
For a function $f$ that satisfies the requirements of Definition~\ref{def:modeling-s-curves},
we define $\wsweet$ and $\wsat$ as follows:
\begin{eqnarray}
\label{eq:YcurveTrans}
y(w) &=& 
    \ln(\frac{1}{2 f(w)} - 1)
\\
\label{eq:wsweet}
\wsweet &=& 
\max \left\{ w \in \mathbb{N} \middle| y''(w) \geq \Gamma |y'(w)| \right\}
\\
\label{eq:wsat}
\wsat &=& 
\max \left\{ w \in \mathbb{N} \;\middle|\; 
f(w) < 0.25 \right\}
\end{eqnarray}
We introduce $\Gamma$ as the knob to control accuracy versus cost. $\Gamma$ is a hyperparameter with a default value $\Gamma=1$. The idea behind the definition of $\wsweet$ lies in the transformed curve $y(w)$ that we define in Equation~\ref{eq:YcurveTrans} and illustrate in Figure~\ref{fig:ScurveTransformedData} for our running example of a surface code with distance 7.
We call this curve a Y-curve.

If we plug our S-curve model in Definition~\ref{def:our-s-curve-model} into Equation~\ref{eq:YcurveTrans}, we get, for $t < w$:

\begin{eqnarray*}
    y_t[\mu,\alpha,\beta](w) 
    &=&
    \ln(\frac{1}{2 f_t[\mu,\alpha,\beta](w)-1}) \\
    &=&
    -\frac{1}{\alpha} w  + \frac{\mu}{\alpha} + \frac{\beta}{\sqrt{(w-t)} } 
    \label{eq:linear}
\end{eqnarray*}

The function $y_t$ has linear and non-linear components. We provide a physical interpretation for this behavior. The linear term $-\frac{1}{\alpha} w+ \frac{\mu}{\alpha}$, captures the overall degradation of error-correcting performance as the error weight $w$ increases. The non-linear term, $\frac{\beta}{\sqrt{w-t}}$, models the increased ability of the QEC circuit to correct low-weight errors.

The first-order derivative of $y_t$ is
\begin{eqnarray*}
y_t'(w) &=& -\frac{1}{\alpha} - \frac{\beta}{2} \frac{1}{(w-t)^{\frac{3}{2}}}, \;\; \lim_{w \rightarrow \infty} y_t(w) = -\frac{1}{\alpha}
\end{eqnarray*}
In the formula above for $y_t'(w)$, the first term says that the almost-linear portion of the curve has slope $-\frac{1}{\alpha}$.
The second term is non-linear but diminishes quickly as $w$ increases. 
The second-order derivative of $y_t$ is:

\begin{equation}
    y_t''(w) = \frac{3\beta}{4} \frac{1}{(w-t)^{\frac{5}{2}}}, \quad \lim_{w \rightarrow \infty} y_t''(w) = 0
\end{equation}

We are looking for a sweet spot where the second order derivative is comparable with the absolute value of the first order derivative, that is, well before the second-order derivative gets close to 0.
This leads us to Equation~\ref{eq:wsweet}, which defines $\wsweet$ so that we have 
\begin{equation*}
      y''(\wsweet) \approx \Gamma |y'(\wsweet)|
   \label{eq:SweetSpot}
\end{equation*}
The idea behind the definition of $\wsat$ is to ensure that we will focus on the weights that are in the left half of the S-curve.  Ideally, we would define $\wsat$ as the inflection point where the steep slope ends and the S-curve starts to flatten out.  Mathematically, this means that we want $\wsat$ to satisfy $f''(\wsat) \approx 0$. Instead, Equation~\ref{eq:wsat} is a more practical and approximate definition of the inflection point.  From Definition~\ref{def:modeling-s-curves}, we know that $f(w)$ increases from 0 to 0.5, and from Figure~\ref{fig:allcodes} we see that the inflection point is roughly when $f(w)$ is half way between 0 and 0.5, which is 0.25.  So, we define $\wsat$ so that it satisfies 
\begin{equation*}
f(\wsat) \approx 0.25
\end{equation*}

For example, for the surface code with distance 7 (discussed in Sections~\ref{sec:introduction}--\ref{sec:modeling-qec-test-data}),
if we use our S-curve model in Definition~\ref{def:our-s-curve-model}, fitted to the QEC-test data as described in Section~\ref{sec:modeling-qec-test-data}, we get $\wsweet = 12$ and $\wsat = 65$. 
We will test only high-weight samples, which for this example are the ones with weights between 12 and 65.

\section{Testing Samples with High Weights}
\label{sec:testing-samples-with-high-weights}

Our approach focuses on testing samples with high weights and then extrapolating to low weights.
The idea is to test samples with weights between $\wsweet$ and $\wsat$, and obtain values for the model parameters in the S-curve model.
The challenge is that this is circular: we want to use $\wsweet$ and $\wsat$ to determine $f$, but we need $f$ to determine $\wsweet$ and $\wsat$.
In this section, we explain how we break this cycle and successfully test within a limited time budget.

\subsection{Our Approach}

\begin{algorithm}[t!]
\hrule height 1pt
\caption{\texorpdfstring{\textsc{AdapSam}}{AdapSam} --- Stratified and adaptive subspace sampling algorithm for S-curve fitting}
\label{alg:adaptive-sampling}
\hrule height 0.5pt
\begin{algorithmic}[1]
\Require Total sampling budget $\maxshots$, Y-curve model template $y_t[\vec{v}]$, minimum logical errors per subspace $N_{LE}$, hyperparameter $\Gamma$.
\Ensure Fitted parameter $\vec{v}$ of Y-curve model.
\Procedure{AdapSam}{$\maxshots, N_{LE}, \Gamma$}

    \State \textbf{Stage 1: Binary-search stage}
    \State $\werr \gets$ smallest $w$ with observed logical errors with \mbox{1,000} samples (binary search)
    \State $w_{\text{sat}} \gets$ largest $w$ with $\hat{P}_L(w) \leq 0.25$ 
    \Statex \hspace*{1cm} (binary search)

    \Statex
    \State \textbf{Stage 2: Initial-sampling stage}
    \State $\mathcal{W} \gets$ 5 points in $[\werr, w_{\text{sat}}]$, uniformly
    \ForAll{$w \in \mathcal{W}$}
        \State Sample at $w$ until 
        \Statex \hspace*{1.6cm} $N_{LE}$ logical errors are collected
    \EndFor
    \State $w_{\text{front}} \gets \min(\mathcal{W})$

    \Statex %
    \State \textbf{Stage 3: Iterative stage.}
    \While{$|\mathcal{W}| < 10 \wedge \mbox{\#samples}\ \leq \maxshots$}
        \State Fit Y-curve: $\vec{v} \gets \textsc{FitYCurve}(\text{data})$
        \State $\hat{w}_{\text{sweet}} \gets \max \big\{ w \;\big|\; y''(w) \geq \Gamma |y'(w)| \big\}$ %
        \If{$w_{\text{front}} \leq \hat{w}_{\text{sweet}}$} \textbf{break}
        \EndIf
        \State $\Delta w \gets \lfloor (\werr - \hat{w}_{\text{sweet}}) / 5 \rfloor$
        \State $w_{\text{new}} \gets \max(w_{\text{front}} - \Delta w, \hat{w}_{\text{sweet}})$
        \State Sample at $w_{\text{new}}$ until 
        \Statex \hspace*{1.5cm} $N_{LE}$ logical errors collected
        \State $\mathcal{W} \gets \mathcal{W} \cup \{w_{\text{new}}\}$; \; $w_{\text{front}} \gets w_{\text{new}}$
    \EndWhile
    \Statex %
    \State Fit final Y-curve.
    \State \Return Fitted parameters $\vec{v}$.
\EndProcedure
\end{algorithmic}
\hrule height 1pt
\end{algorithm}

\begin{figure}[ht!]
    \centering
    \includegraphics[width=0.48\textwidth]{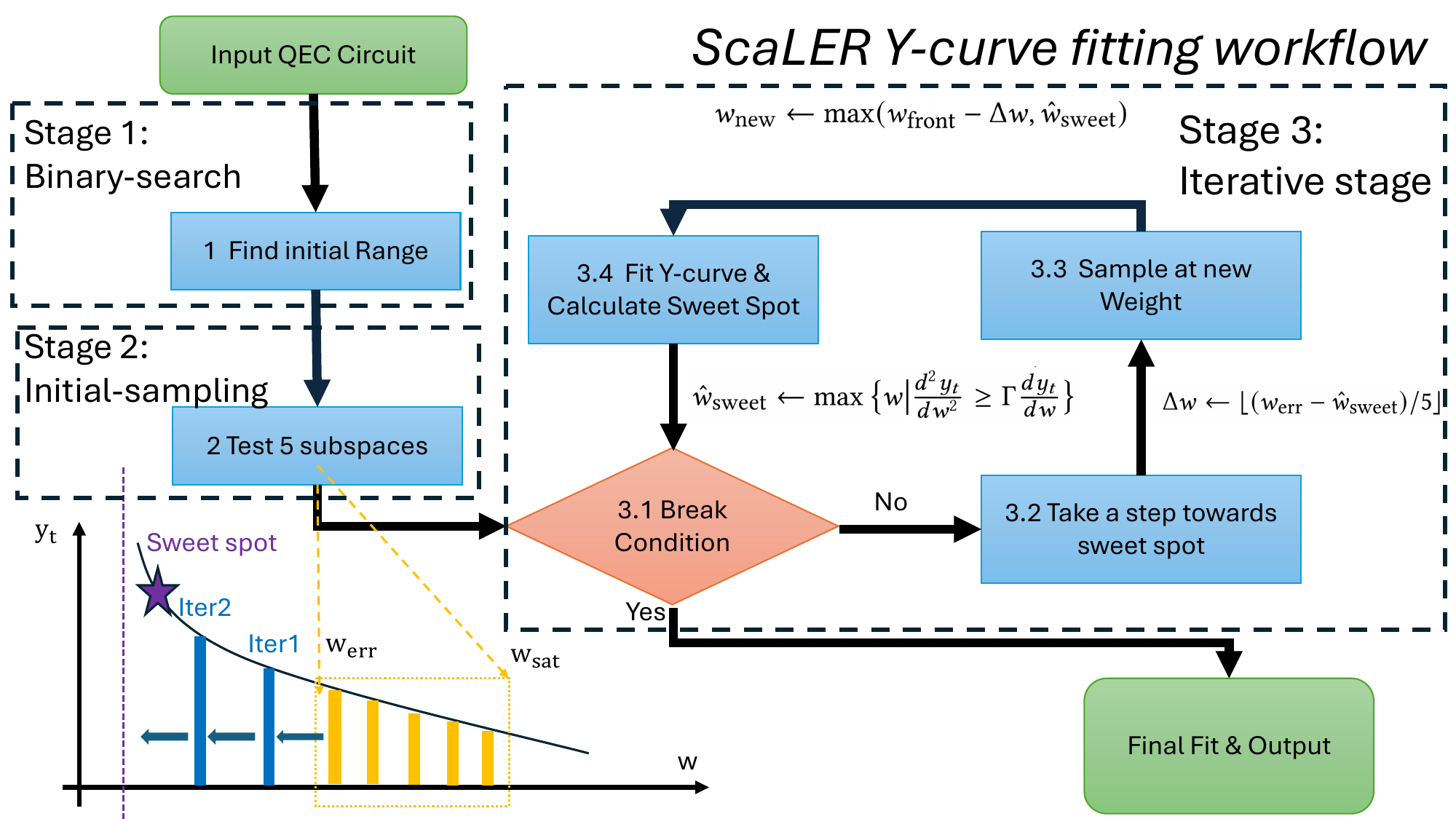}
    \caption{Diagram for the adaptive Y-curve fitting approach used in \ourtool.}
    \label{fig:Diagram}
\end{figure}

Our approach is described in Algorithm \ref{alg:adaptive-sampling} and illustrated in Figure \ref{fig:Diagram}. 
The idea is to test samples with lower and lower weights, proceeding in three stages.

For a weight $w$, we will refer to the set of samples with weight $w$ as a subspace.

\paragraph{The binary-search stage.}
In the first stage, we use binary search to determine both $\wsat$ and also a weight $\werr$ that satisfies the following:
\begin{equation*}
P^{\werr}_L \approx 0.001
\end{equation*}
For QEC-implementations with high distances, we have $\wsweet < \werr < \wsat$.  We use \mbox{1,000} samples for each weight, so $\werr$ is a weight for which testing \mbox{1,000} samples gives a single logical error, while $\wsat$ is a weight for which testing \mbox{1,000} samples gives less than 250 errors.

\paragraph{The initial-sampling stage.}

In the second stage, we test in 5 subspaces that are evenly distributed between $\werr$ and $\wsat$, and we fit a Y-curve to the data. From the Y-curve we estimate $\wsweet$.

\paragraph{The iterative stage.}

In the third stage, we test in 5 subspaces that are evenly distributed between $\wsweet$ and $\werr$.
Each time we finish a subspace, we update the Y-curve and update the estimate of $\wsweet$. 
Testing concludes either when 5 subspaces have been tested or when the time budget is exhausted.

\paragraph{Stopping criterion per subspace.}
We generate random samples in a given subspace until more than $N_{\text{LE}}$ logical errors are observed. The default is $N_{\text{LE}} = 30$. 

\begin{table}[t]
\centering
\small
\resizebox{\columnwidth}{!}{%
\begin{tabular}{|c|p{6.5cm}|}
\hline
\textbf{Symbol} & \textbf{Description} \\
\hline
$S_{\max}$ %
& Total sampling budget (equivalent to a time budget) \\
\hline
$N_{LE}$ 
& Required logical errors per subspace (set to $30$ in this paper) \\
\hline
$\Gamma$ 
& Sweet-spot curvature threshold (Eq.~\ref{eq:SweetSpot}) (set to $1$ in this paper) \\
\hline
\end{tabular}
}
\caption{Hyperparameters in \textsc{AdaptiveSampling}.}
\label{tab:Hyperparameter}
\end{table}

\subsection{Example}
\label{sec:example-of-how-our-approach-works}
\begin{figure}[h!]
\centering
\begin{minipage}{0.47\textwidth}
    \centering
    \includegraphics[width=\linewidth]{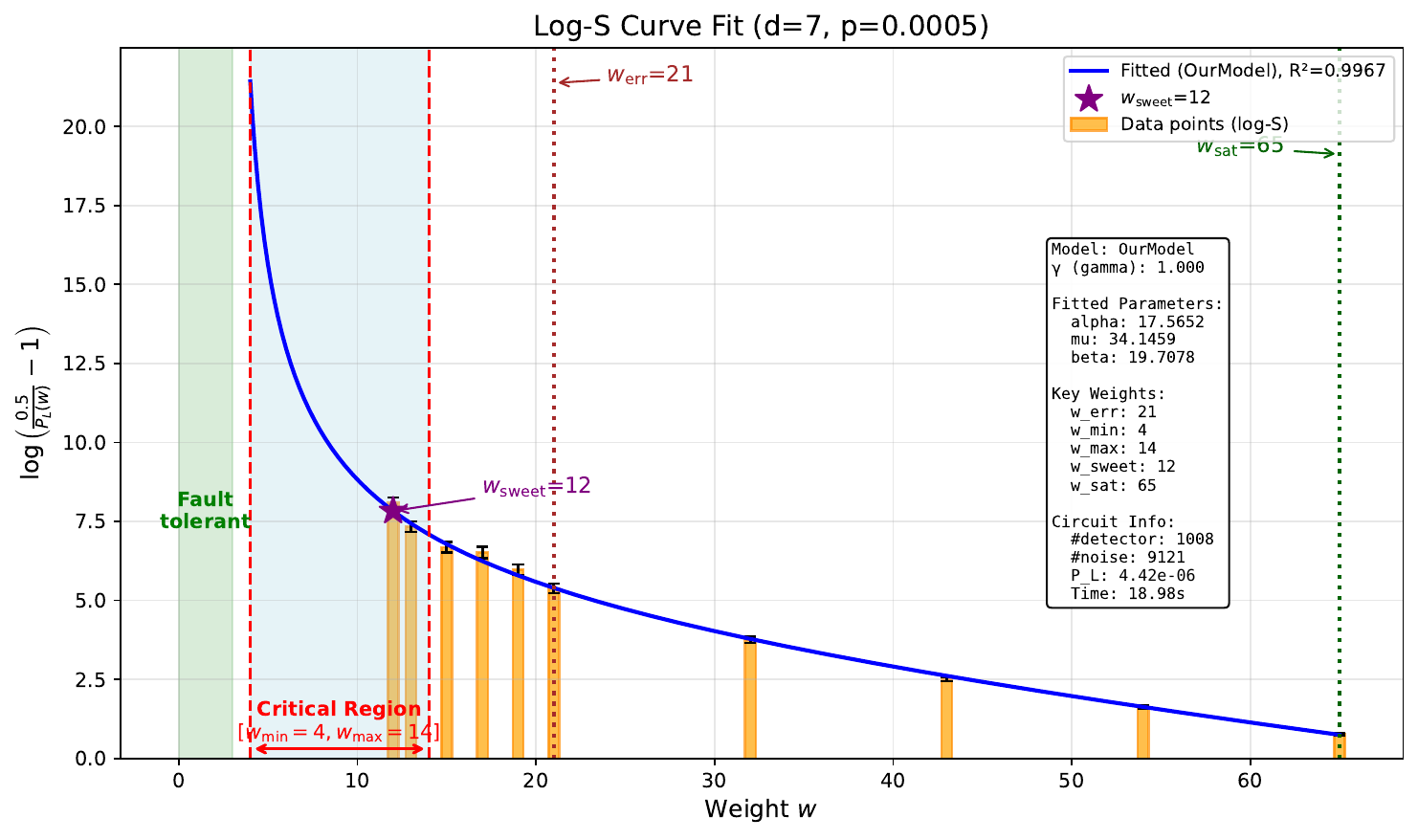}
\end{minipage}
\caption{Examples of fitted curves for the surface code with distance $7$ when $p=0.0005$.} 
\label{fig:surface7Example}
\end{figure}

For the surface code with distance 7 (discussed in Sections~\ref{sec:introduction}--\ref{sec:separating-low-and-high-weights}), Algorithm~\ref{alg:adaptive-sampling} proceeds as follows. 

\paragraph{Compilation.}

A distance-$d$ surface code uses $2d^2-1$ qubits. Therefore, our distance-7 surface code involves a $97$-qubit circuit that contains $1008$ detectors. 
We repeat each stabilizer-measurement round $3d=21$ times to preserve fault tolerance in the circuit-level noise model 
\cite{shor1996fault}. 

In the SID noise model, we consider all fault locations in the circuit. This circuit has $n=\mbox{9,121}$ locations, and we compile the corresponding quantum error propagation graph (QEPG), stored as a $\mbox{9,121} \times 1008$ matrix that represents the propagation of Pauli errors to the detectors.

\paragraph{The binary-search stage.}
The range of binary search is $1\leq w \leq \mbox{9,121}$. For each subspace with weight $w$ that we encounter in the search, we generate $\mbox{1,000}$ random samples.  The outcome is that we find $\werr=21$ and $\wsat=65$.

\begin{table}[t]
\centering
\small
\begin{tabular}{rrrr}
\toprule
\textbf{Weight} 
& $\boldsymbol{N_w^{\mathrm{LE}}}$ 
& $\boldsymbol{N_w^{\mathrm{Test}}}$ 
& \multicolumn{1}{c}{$\boldsymbol{\hat{P}^w_L}$} \\
\midrule
$\wsweet$ = 12 &   32 & 207{,}499 & $1.54 \times 10^{-4}$ \\
13 &   35 & 107{,}499 & $3.26 \times 10^{-4}$ \\
15 &   36 &  57{,}499 & $6.26 \times 10^{-4}$ \\
17 &   30 &  40{,}833 & $7.35 \times 10^{-4}$ \\
19 &   36 &  28{,}333 & $1.27 \times 10^{-3}$ \\
$\werr$ = 21 &   42 &  18{,}333 & $2.29 \times 10^{-3}$ \\
32 &  114 &  10{,}000 & $1.14 \times 10^{-2}$ \\
43 &  377 &  10{,}000 & $3.77 \times 10^{-2}$ \\
54 &  821 &  10{,}000 & $8.21 \times 10^{-2}$ \\
$\wsat$ = 65 & 1579 &  10{,}000 & $1.58 \times 10^{-1}$ \\
\bottomrule
\end{tabular}
\caption{Logical error statistics per error weight $w$. Here $N_w^{\mathrm{LE}}$ is the number of observed logical errors, $N_w^{\mathrm{Test}}$ is the total number of trials, and $\hat{P}^w_L = N_w^{\mathrm{LE}} / N_w^{\mathrm{Test}}$ is the estimated logical error rate.}
\label{tab:le_sample_data}
\end{table}

\paragraph{The initial-sampling stage.}

We test in the five subspaces with weights 
\begin{equation*}
\werr = 21 \leq 32 \leq 43 \leq 54 \leq 65 = \wsat
\end{equation*}
Table~\ref{tab:le_sample_data} shows the results.
For example, for the subspace with weight 21, we test 18,333 samples and observe 42 logical errors. 
The outcome is that we estimate $\wsweet =13$.

\paragraph{The iterative stage.}

We tested a succession of subspaces and kept updating our estimate of $\wsweet$.  Eventually, we estimated $\wsweet = 12$, after testing the five subspaces with weights
\begin{equation*}
\wsweet = 12 \leq 13 \leq 15 \leq 17 \leq 19 < \werr
\end{equation*}
Table~\ref{tab:le_sample_data} shows the results.

We fit a Y-curve to the data.  For our S-curve model, we estimate the parameters as 
$\hat{\alpha}=17.57$, $\hat{\mu}=34.14$, and $\hat{\beta}=19.71$.

\paragraph{Comparison with Monte Carlo sampling using \stim.}

Figure~\ref{fig:bar-diagram-number-of-samples-for-each weight} presents Monte Carlo data generated using \stim from testing the distance-7 surface code, where we instrumented \stim with additional logging hooks and ran the simulator for a two-hour time budget and observed $503$ logical errors.

Monte Carlo sampling using \stim uses $83{,}000{,}000$ samples to observe $503$ logical errors. In contrast, \ourtool uses approximately $5 \times 10^{5}$ samples in total, that is, only $0.6\%$ of the Monte Carlo baseline's sample count. Table~\ref{tab:le_sample_data} shows that most of the sampling budget of \ourtool is concentrated on the two smallest weights near the sweet spot, namely 12 and 13.  For these weights, the subspace logical error rate is low and therefore hard to estimate. However, Figure~\ref{fig:bar-diagram-number-of-samples-for-each weight} shows that Monte Carlo sampling using \stim primarily tests subspaces with weights between 1 and 10, which are significantly more difficult to test than subspaces with the higher weights 12 and 13.

\section{Our Algorithm}
\label{sec:our-algorithm}

In this section, we present our overall algorithm \ourtool (which abbreviates Scalable Logical Error Rate testing). 
The idea is to first test samples with high weights and derive an S-curve that models the test data, and then use the S-curve to estimate the logical error rate.

\subsection{Pseudo-code for Our Algorithm}

\begin{algorithm}[t]
\hrule height 1pt
\caption{\textsc{{\ourtool}} — Stratified fault injection and S-Curve Fitting for Logical-Error-Rate Estimation}
\label{alg:ScaLER}
\hrule height 0.5pt
\begin{algorithmic}[1]
\Require Circuit $\mathcal{C}$ with circuit-level code distance $d$, physical error rate $p$, total sample budget $\maxshots$, and total error budget $\maxerrors$, sweet-spot curvature threshold $\Gamma$. 
\Ensure $\hat{P}_L$
\Procedure{\ourtool}{$\mathcal{C}, d, p, \maxshots, \maxerrors, \Gamma$}
    \State $t \gets \frac{d-1}{2}$
    \State $\vec{v} \gets \Call{AdapSam}{\maxshots,\maxerrors,\Gamma}$
    \State $n \gets$ number of locations in $\mathcal{C}$
    \State $\displaystyle\hat{P}_L=\sum_{w=t+1}^{n}
             f_t[\vec{v}](w)\ \cdot$
    $\binom{n}{w} p^w (1-p)^{n-w}$
    \State \Return $\hat{P}_L$
\EndProcedure
\end{algorithmic}
\hrule height 1pt
\end{algorithm}

Algorithm \ref{alg:ScaLER} presents pseudo-code for our algorithm.
The input consists of a QEC circuit $\mathcal{C}$ with a circuit-level code distance $d$, the physical error rate $p$, the total sample budget $\maxshots$, the maximum number of logical errors $\maxerrors$ to sample in one subspace $N$, and the sweet-spot curvature threshold $\Gamma$.

In line 2, we calculate the maximum number of errors that $C$ is designed to correct.

In line 3, we run our testing procedure from Section~\ref{sec:testing-samples-with-high-weights}, which produces the S-curve parameters. 

In line 4, we calculate the number of locations in the circuit $\mathcal{C}$.

In line 5, we estimate the logical error rate by applying Equation~\ref{eq:stratified-estimate-of-logical-error-rate}, with $P^w_L$ replaced by the model $f_t[\vec{v}](w)$, summing over all weights $w>t$ (because a sample with weight at most $t$ does not cause a logical error).

Finally, in line 6, we return the estimated $\hat{P}_L$.

\subsection{Example}

For our running example of a surface code with distance 7 (discussed in Sections~\ref{sec:introduction}--\ref{sec:testing-samples-with-high-weights}), we have a circuit with $n = \mbox{9,121}$ locations and we have the physical error rate $p = 0.0005$.
The expected number of injected errors is $\mu = np = \mbox{9,121} \cdot 5 \cdot  10^{-4} \approx 4$, with the standard deviation
$\sigma = \sqrt{np(1-p)} \approx \sqrt{\mbox{9,121} \cdot 5 \times 10^{-4}(1-5 \cdot 10^{-4})} \approx 2$.
Using a coverage factor $k=5$, we define the critical error-weight region as
$[\mu - k\sigma,\; \mu + k\sigma] \approx [4 - 5\cdot 2,\; 4 + 5\cdot 2]$.
Since the error weight below $3$ is in the fault-tolerant zone, we set the lower bound to $4$. The critical region (\textcolor{blue}{blue} color in Figure \ref{fig:surface7Example}) is therefore $[w_{\text{min}}=4,w_{\text{max}}=14]$.
Lines 2--3 of Algorithm~\ref{alg:ScaLER} produce the S-curve $f_t[17.57,34.14,19.71](w)$, after which lines 4--5 produce the following estimate of the logical error rate:
\begin{align}
\hat{P}_L
&= \sum_{w=4}^{\mbox{9,121}} f_t[17.57,34.14,19.71](w)\,\binom{\mbox{9,121}}{w}
\notag\\
&\quad\cdot (5\times 10^{-4})^{w}\,(1-5\times 10^{-4})^{n-w}
\notag\\
&\approx 4.42\times 10^{-6}.\label{eq:surface-logical-error-estimate}
\end{align}

The estimated logical error rate for Monte Carlo sampling using \stim is $\frac{503}{83{,}000{,}000}\approx 6.06 \times 10^{-6}$.  \ourtool is off by $27 \%$, compared to Monte Carlo sampling using \stim in this case. We will further explain the cause of the inaccuracy and how to improve in Section \S \ref{sec:RQ2}.

\section{Implementation}
\label{sec:implementation}

\begin{figure}[ht!]
    \centering
    \includegraphics[width=0.45\textwidth]{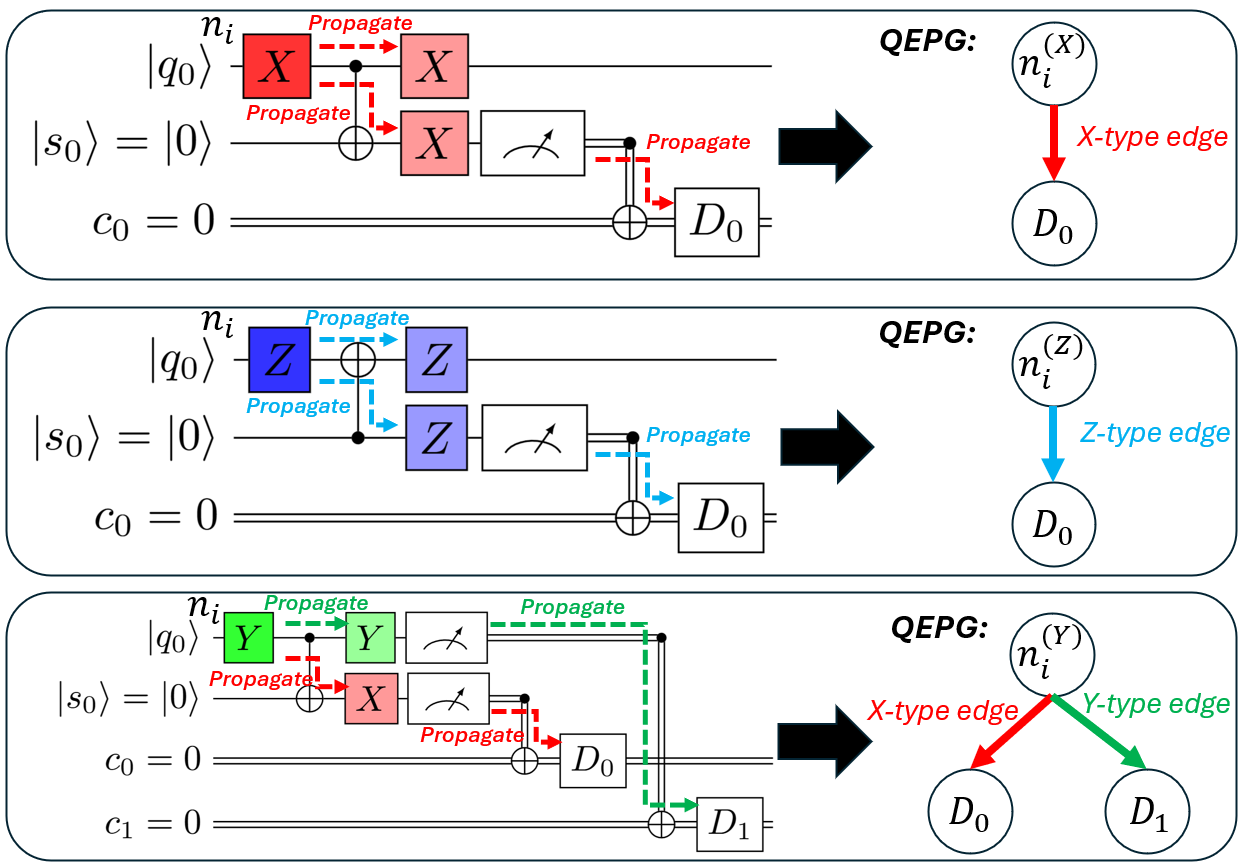}
    \caption{Construction of a QEPG graph in \ourtool.}
    \label{fig:prop}
\end{figure}

We implemented \ourtool in C++ and Python.
The C++ part encodes the error model and compiles each sample in a manner similar to \stim \cite{gidney2021stim}.
In particular, it compiles a circuit 
to a binary graph that we call a quantum error propagation graph (QEPG) in which every vertex is a Pauli gate and each edge denotes a type of propagation, as illustrated in Figure \ref{fig:prop}.
In addition, it uses binary vector addition to simulate Pauli error propagation, uses
Boost’s dynamic\_bitset \cite{boost_dynamic_bitset} to represent and update the QEPG graph, and accelerates the computation by parallelizing across threads with OpenMP \cite{dagum1998openmp}. 

The Python part implements Algorithms~\ref{alg:adaptive-sampling}--\ref{alg:ScaLER}, exposing the generated samples to Python as NumPy arrays using pybind11 \cite{pybind11_docs} and using the Python scipy library \cite{2020SciPy-NMeth} to fit the Y-curve model and calculate the logical error rate.

\paragraph{Access.}

We have released \ourtool as an \href{https://github.com/yezhuoyang/ScaLERQEC}{open-source project} on GitHub, including a README file that explains how to install our tool, run it, and reproduce all the results in this paper.

\section{Evaluation}
\label{sec:evaluation}

\begin{table*}[!t]
\centering
\small
\renewcommand{\arraystretch}{0.92}
\setlength{\tabcolsep}{3.5pt}

\resizebox{\textwidth}{!}{%
\begin{tabular}{|c|c|c|c|c|c|c|c|}
\hline
\multicolumn{8}{|c|}{\textbf{$p=0.0005$}} \\
\hline
\textbf{Code} 
& \textbf{d} 
& \textbf{MC (\stim) LER}
& \textbf{MC (\stim) Sampled LE}
& \textbf{MC (\stim) (s)}
& \textbf{\ourtool LER} 
& \textbf{\ourtool (s)} 
& \textbf{Rel.~error} \\
\hline
BBCode$[[72,12,6]]$   & 6  & $(3.76 \pm 0.12) \times 10^{-4}$ & 102 & 562  & $(4.41 \pm 0.43) \times 10^{-4}$ & 559  & $-17.3\%$ \\
BBCode$[[90,8,10]]$   & 10 & $(8.68 \pm 0.36) \times 10^{-5}$ & 100 & 6839 & $(1.01 \pm 0.04) \times 10^{-4}$ & 5447 & $-16.4\%$ \\
BBCode$[[108,8,8]]$   & 8  & $(5.89 \pm 1.29) \times 10^{-6}$ & 8 & 7200 & $(4.20 \pm 0.70) \times 10^{-6}$ & 6044 & $(28.7\%)$ \\
BBCode$[[144,12,12]]$ & 12 & 0 & 0 & 7200 & $(4.16 \pm 0.86) \times 10^{-6}$ & 7065 & --- \\
\hline
\multicolumn{8}{|c|}{\textbf{$p=0.0001$}} \\
\hline
\textbf{Code} 
& \textbf{d} 
& \textbf{MC (\stim) LER}
& \textbf{MC (\stim) Sampled LE}
& \textbf{MC (\stim) (s)}
& \textbf{\ourtool LER} 
& \textbf{\ourtool (s)} 
& \textbf{Rel.~error} \\
\hline
BBCode$[[72,12,6]]$   & 6  & $(2.47 \pm 0.3) \times 10^{-6}$ & 19 & 7200  & $(1.37 \pm 0.72) \times 10^{-6}$ &  2436   & $(-44.5\%)$ \\
BBCode$[[90,8,10]]$   & 10 & $(6.59 \pm 2.7) \times 10^{-7}$ & 2 & 7200 & $(2.63 \pm 1.0) \times 10^{-7}$ & 6686 & $(-60.1\%)$\\
BBCode$[[108,8,8]]$   & 8  & 0 & 0 & 7200 & $(6.45 \pm 0.98) \times 10^{-11}$ & 6795 & --- \\
BBCode$[[144,12,12]]$ & 12 & 0 & 0 & 7200 & $(5.35 \pm 0.13) \times 10^{-12}$ & 6209 & --- \\
\hline
\end{tabular}
}

\vspace{6pt} %

\resizebox{\textwidth}{!}{%
\begin{tabular}{|c|c|c|c|c|c|c|c|}
\hline
\textbf{Code} 
& \textbf{d} 
& \textbf{MC (\stim) LER}
& \textbf{MC (\stim) Sampled LE}
& \textbf{MC (\stim) (s)}
& \textbf{\ourtool LER} 
& \textbf{\ourtool (s)} 
& \textbf{Rel.~error} \\
\hline
& 3  & $(5.77 \pm 0.19) \times 10^{-4}$ & $184$ & 0.34 & $(5.81 \pm 0.64) \times 10^{-4}$ & 0.83   & $-0.7\%$ \\
& 5  & $(6.41 \pm 0.35) \times 10^{-5}$ & $115$ & 8.7  & $(4.51 \pm 0.43) \times 10^{-5}$ & 4.6    & $29.6\%$ \\
Surface
& 7  & $(5.95 \pm 0.66) \times 10^{-6}$ & $101$  & 216  & $(3.75 \pm 0.28) \times 10^{-6}$ & 20     & $37.0\%$ \\
& 9  & $(3.59 \pm 0.59) \times 10^{-7}$ & $36$   & 3116 & $(3.54 \pm 0.44) \times 10^{-7}$ & 95     & $1.4\%$ \\
& 11 & $(2.22 \pm 1.26) \times 10^{-8}$ & $3$    & 7213 & $(2.92 \pm 0.25) \times 10^{-8}$ & 3337   & $(31.5\%)$ \\
& 13 & 0 & 0 & 7200 & $(2.45 \pm 0.17) \times 10^{-9}$  & 6287 & --- \\
& 15 & 0 & 0 & 7200 & $(1.65 \pm 0.29) \times 10^{-10}$ & 7200 & --- \\
& 17 & 0 & 0 & 7200 & $(1.51 \pm 0.08) \times 10^{-11}$ & 7200 & --- \\
\hline
& 3  & $(7.05 \pm 0.35) \times 10^{-4}$ & $159$ & 0.16 & $(7.51 \pm 0.22) \times 10^{-4}$ & 0.85   & $-6.5\%$ \\
& 5  & $(8.32 \pm 0.47) \times 10^{-5}$ & $125$ & 5.9  & $(4.99 \pm 0.27) \times 10^{-5}$ & 4.7    & $40.0\%$ \\
Toric
& 7  & $(6.82 \pm 0.54) \times 10^{-6}$ & $101$  & 162  & $(4.50 \pm 0.22) \times 10^{-6}$ & 20     & $34.0\%$ \\
& 9  & $(4.56 \pm 0.38) \times 10^{-7}$ & $100$  & 1998 & $(4.24 \pm 0.29) \times 10^{-7}$ & 125    & $7.0\%$ \\
& 11 & $(3.53 \pm 1.6) \times 10^{-8}$  & $5$    & 7200 & $(3.63 \pm 0.15) \times 10^{-8}$ & 2282   & $(2.83\%)$ \\
& 13 & 0 & 0 & 7200 & $(2.46 \pm 0.22) \times 10^{-9}$  & 5483 & --- \\
& 15 & 0 & 0 & 7200 & $(2.28 \pm 0.23) \times 10^{-10}$ & 7200 & --- \\
\hline
\end{tabular}
}

\caption{Comparison of LER estimation: unconditional Monte Carlo sampling using \stim{} (MC) vs.\ \ourtool. (Top) BB codes at $p=0.0005$ and $p=0.0001$.
(Bottom) Surface and Toric codes at $p=0.0005$. We use a 2-hour time budget; BB is repeated 3 times and Surface/Toric is repeated 5 times. For larger instances, Monte Carlo sampling using \stim observes no logical errors within budget, while \ourtool still produces an estimate.}
\label{tab:ler_comparison_all_packed}
\end{table*}

\begin{figure*}[t]
    \centering
    \includegraphics[width=\textwidth]{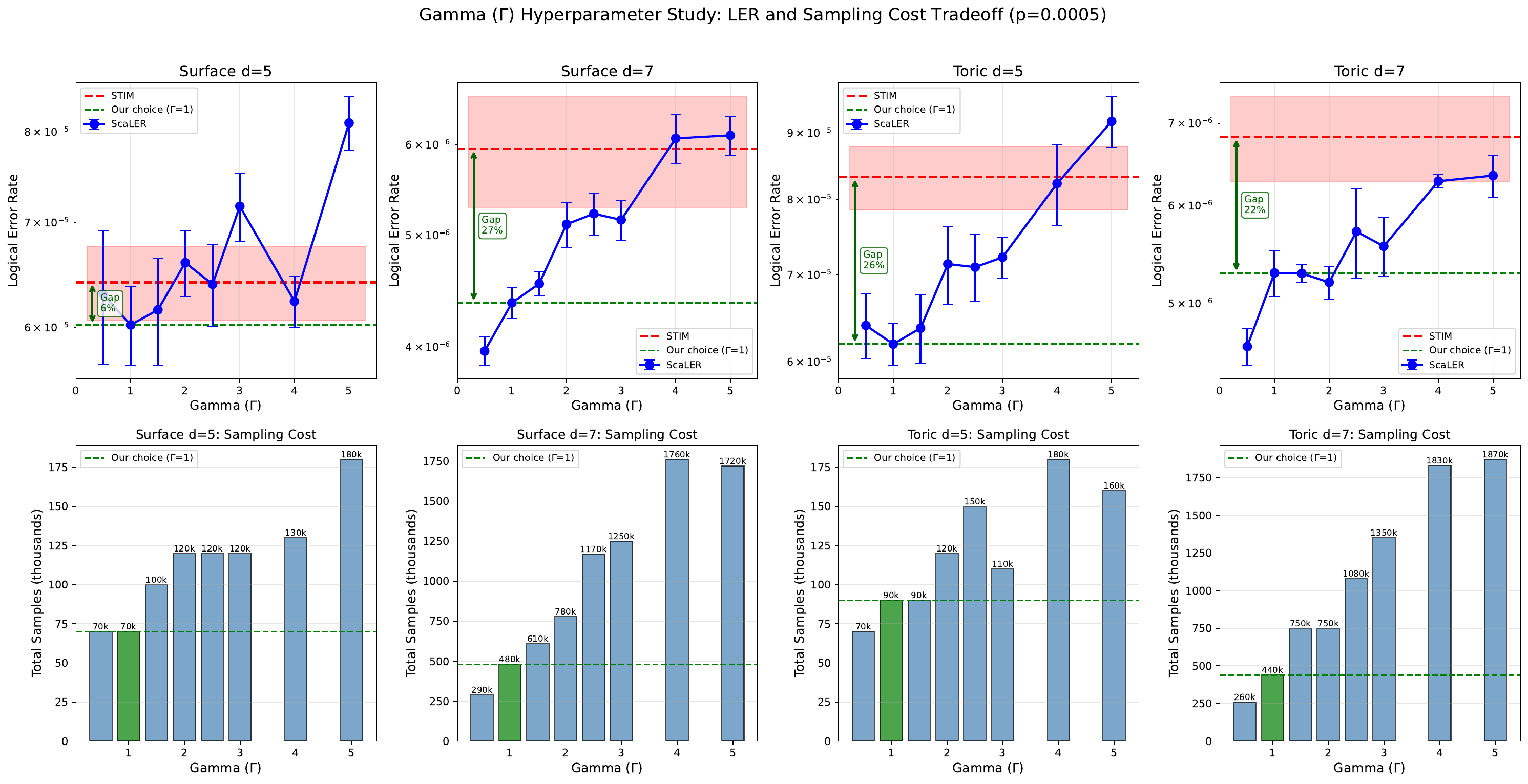}
    \caption{We measure the accuracy-cost tradeoff of different Hyperparameter $\Gamma$ selection of our algorithm at small scale when code distance is $5$ and $7$. Above: We plot the logical error rate estimated by \ourtool at different $\Gamma$ together with output of Monte Carlo sampling using \stim. Below: We plot the total sampling cost of different Hyperparameters. When $\Gamma$ increases, \ourtool gets logical error rate closer to Monte Carlo sampling using \stim, with a large sampling cost. At $\Gamma=1$, we save a tremendous amount of sampling cost while has an inaccuracy around $20\%$, which is acceptable.}
    \label{fig:Tradeoff}
\end{figure*}

Our experimental evaluation answers three research questions (RQs) about our approach. The questions and our answers are as follows.

\begin{itemize}

    \item[RQ1] \textbf{Does \ourtool scale better than unconditional Monte Carlo sampling using \stim?} Yes, at the physical error rate 0.0005 with a time budget of 2 hours, unconditional Monte Carlo sampling using \stim cannot sample any logical error with a two hour time budget for distance $13$.  \ourtool scaled to distance 17.
    
    \item[RQ2] \textbf{Does \ourtool agree with Monte Carlo sampling using \stim?} Yes, more than half of the results have an inaccuracy below $20\%$. For the remaining cases, we show that the accuracy can be improved by changing the hyperparameter settings.

    \item[RQ3] \textbf{What is the best S-curve model?} We tested both IBM's S-curve model and variants of our S-curve model and found that our model with $\Gamma = 1$ achieves the highest accuracy.

\end{itemize}

In the remainder of this section, we show how our evaluation results support our claims.

\paragraph{Benchmarks.}

We experiment with both standard topological codes and QLDPC codes.
Our benchmarks are three sets of QEC programs from previous work:
the standard surface code and the Toric code with distances up to 17, from the Gidney artifact \cite{gidney2021stim}, and four different Bivariate Bicycle QLDPC codes
\cite{Bravyi_2024} with distances up to 12, from Perlin et al.~\cite{perlin2023qldpc}.

The Surface code, Toric code, and variants are decoded by Pymatching
\cite{higgott2023sparse}, the Bivariate Bicycle Qldpc code is decoded by the Belief Propagation with Ordered Statistics Decoding (BP-OSD) algorithm
\cite{bposd}.

We set the physical error rate at $p = 0.0005$ for most experiments, except where noted explicitly. This rate is a realistic setting because many platforms have a physical error rate for 1-qubit gates that is below $10^{-3}$ \cite{ballance2016high,huang2019fidelity,rol2019fast,jurcevic2021demonstration,foxen2020demonstrating,wu2021strong}.

\paragraph{Experimental Setup.}
\label{sec:experimental-setup}

We conducted all experiments on a Windows desktop with an Alienware Aurora R16 platform, equipped with a 13th Gen Intel Core i9-13900F processor (24 cores, 32 threads) and 32\,GB of RAM. We set a time budget of $2$ hours for each experiment.  We repeated each experiment $5$ times, which allowed us to compute the average value and the standard deviation. 
For each experiment, we record the number of samples, the execution time, and the logical error rate. %

\paragraph{Stopping criterion for Monte Carlo sampling using \stim}
Following previous work, we run Monte Carlo sampling using \stim until it observes $100$ logical errors or exhausts the 2-hour time budget \cite{Gidney2021faulttolerant}. The threshold of 100 logical errors ensures that the resulting statistical uncertainty, measured by relative standard deviation (standard deviation divided by mean estimated value) is less than $10\%$ \cite{wasserman2004all}.  \ourtool matches this.

We compute the relative accuracy of the logical error rate estimated by \ourtool versus Monte Carlo sampling using \stim for all configurations in which Monte Carlo sampling using \stim observes at least $100$ logical errors, and thus its estimated logical error rate has high statistical confidence and can be considered as ground truth.

\subsection{RQ1: Scalability Comparison}
\label{sec:RQ1}

Table~\ref{tab:ler_comparison_all_packed} shows the results for all our benchmarks.

\paragraph{Result for Surface code and Toric code.}

We evaluated the logical error rates for both the Surface and Toric codes with code distances up to $d=17$. For distance $d\leq 10$, Monte Carlo sampling using \stim successfully sampled sufficient logical errors within time budget, producing estimates with small standard deviations and high statistical confidence. However, as the code distance increases, Monte Carlo sampling using \stim struggles to observe rare error events.

In contrast,  \ourtool consistently produces high-confidence logical error rate estimates up to $d=17$ with a 2-hour time budget. 

For example, at the code distance $d=11$ for the Surface code, Monte Carlo sampling using \stim observed only $3$ logical errors after 2 hours, resulting in a standard deviation exceeding $50\%$ of the estimated value. In contrast, \ourtool provided an estimate of $(2.92 \pm 0.25) \times 10^{-8}$ with high confidence in less than $1$ hour.

The results demonstrate that \ourtool is significantly more scalable than unconditional Monte Carlo sampling using \stim. While this Monte Carlo baseline did not sample enough logical failures beyond $d=10$ within the time budget, \ourtool remains robust at higher distances. In particular, \ourtool successfully characterized the Surface code at $d=17$, measuring a logical error rate of $1.51 \times 10^{-11}$, in 2 hours on a desktop. This logical error rate approaches the scale required for real-world quantum applications, such as factoring 2,048-bit RSA integers \cite{gidney2025factor2048bitrsa}.

\paragraph{Result for Bivariate Bicycle code.}

We present the results for the Bivariate Bicycle codes in the top of Table~\ref{tab:ler_comparison_all_packed}. We emphasize that the BP-OSD LDPC decoder is significantly slower than the standard Pymatching decoder, making simulation computationally demanding even at small scales.

The rows in Table~\ref{tab:ler_comparison_all_packed} are ordered by code size. For the first two entries, BBCode [[72,12,6]] and BBCode [[90,8,10]], Monte Carlo sampling using \stim produces logical error rate estimates (on the order of $10^{-4}$) with low standard deviations, successfully sampling over 100 logical errors.

For the larger BBCode [[108,8,8]], Monte Carlo sampling using \stim is unable to sample the 100 logical errors that were targeted within the 2-hour time budget. In fact, only $8$ logical errors were observed, resulting in a dramatic increase in the standard deviation of the estimate.

For the largest code, BBCode [[144,12,12]], Monte Carlo sampling using \stim fails to observe any logical errors and cannot produce a meaningful estimate. In contrast, \ourtool scales effectively, maintaining valid and robust estimates for both cases.

\paragraph{Lower Physical Error Rate.}

To further stress-test the scalability of the tools in the low-error regime, we also did experiments with the lower physical error rate of $p=0.0001$. 
The result in Table \ref{tab:ler_comparison_all_packed} indicates that Monte Carlo sampling using \stim faces significant challenges in this lower error rate regime. Within the time budget of two hours, Monte Carlo sampling using \stim observes only $19$ logical errors for BBCode[[72,12,6]], and just $2$ for BBCode[[90,8,10]], while failing to observe any logical errors for the remaining larger code instances.

In contrast, \ourtool successfully produces logical error rate estimates for all four codes. The results demonstrate that \ourtool exhibits superior scalability compared to Monte Carlo sampling using \stim, particularly as the physical error rate is further reduced.

\paragraph{Stratified Sampling vs.\ Fast Processing.}
How much of the improvement of \ourtool comes from stratified sampling and how much comes from processing each sample efficiently?
We have measured this by implementing plain Monte Carlo sampling ourselves on top of our own QEPG sampler (the same sampler that ScaLER uses, now drawing every fault location independently with probability $p$).  This enabled us to compare our implementation of Monte Carlo sampling with Stim's compiled detector sampler on the same noisy circuits, with the same PyMatching decoder and on the same desktop (Intel Core i5-12400, two OpenMP threads).  For the distance-7 surface code at $p=0.0005$, Stim processes about 49,000 samples per second and our Monte Carlo implementation about 20,000; at distance 11 the numbers are 16,000 and 4,500.  Stim is therefore 2.4 to 3.5 times faster per sample than our implementation of Monte Carlo sampling.  For a two-hour run, this corresponds to 350 million Stim samples against 150 million of ours, at distance 7.  The gain of \ourtool comes from needing far fewer samples. For example, for the distance-7 surface code, \ourtool uses 0.6\% as many samples as Monte Carlo sampling using \stim: \ourtool uses $5\times 10^5$ samples while Monte Carlo sampling using \stim uses $8.3 \times 10^7$ samples.  In summary, \stim is 2.4 to 3.5 times faster than \ourtool at processing each sample but unconditional Monte Carlo sampling using \stim needs many more samples than \ourtool when logical failures are rare.

\begin{figure*}[ht!]
\centering
\begin{minipage}{0.47\textwidth}
    \centering
    \includegraphics[width=\linewidth]{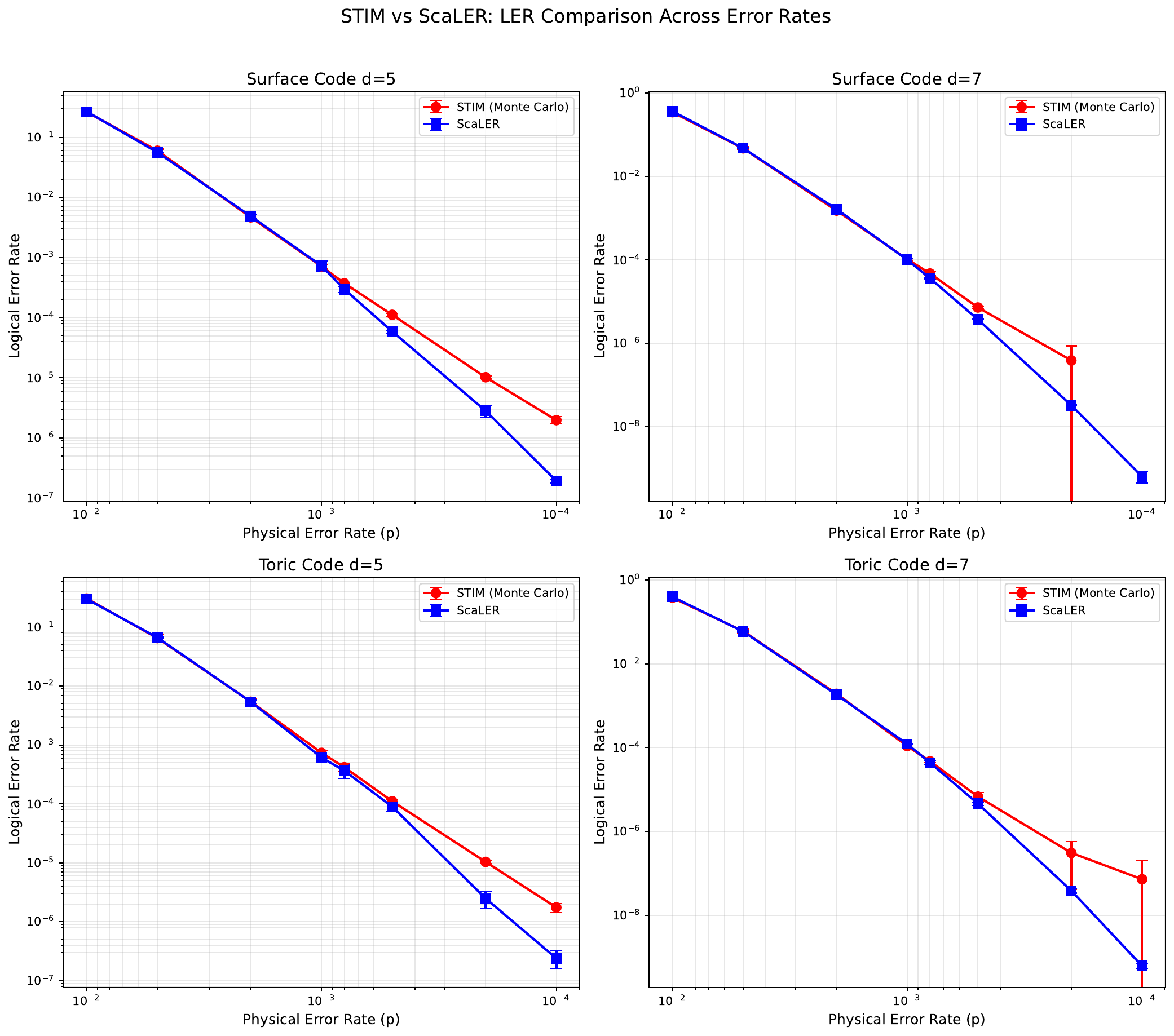}
\end{minipage}
\hfill
\begin{minipage}{0.47\textwidth}
    \centering
    \includegraphics[width=\linewidth]{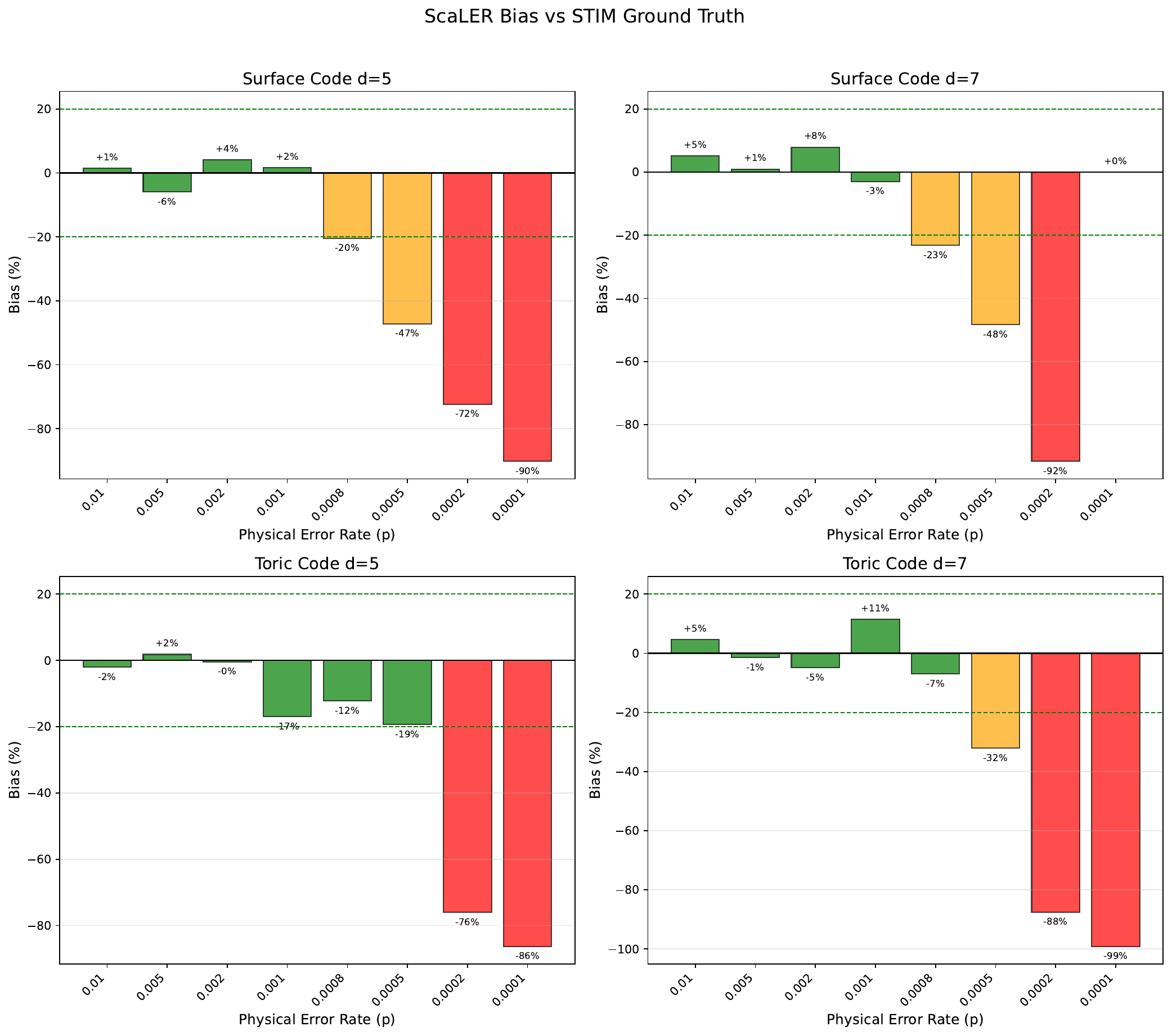}
\end{minipage}
\caption{Study of the relative error of \ourtool compared with Monte Carlo sampling using \stim under different physical error rate ranges from $p=0.01$ to $0.0001$ on Surface code and Toric code. Left figure shows the plot of logical error rate plot of both method, and the right figure shows the relative accuracy of \ourtool compared with Monte Carlo sampling using \stim. First, we observe that both results match well for higher physical error rate, but the accuracy starts to decrease when for lower physical error rate. Second, the results also suggest that the current hyper parameter setting is acceptable if the goal is to achieve a inaccuracy below $20\%$.}
\label{fig:errorrate}
\end{figure*}

\subsection{RQ2: Agreement Check}
\label{sec:RQ2}

To evaluate the accuracy of \ourtool, we compute the relative accuracy of its estimated logical error rates with respect to the baseline results from Monte Carlo sampling using \stim.

\paragraph{Relative error compared with Monte Carlo sampling using \stim.}

As shown in Table \ref{tab:ler_comparison_all_packed}, \ourtool demonstrates strong agreement with Monte Carlo sampling using \stim under our default hyperparameter setting $\Gamma=1$. Indeed, the logical error rate is consistent with Monte Carlo sampling using \stim in order of magnitude with acceptable accuracy. For BBCode [[72,12,6]] and BBCode [[90,8,10]] at $p=0.0005$, where Monte Carlo sampling using \stim observes 100 logical errors within time budget, the inaccuracy of \ourtool compared with Monte Carlo sampling using \stim is below $20\%$.

For the surface code and the toric code with distance $3$ and $9$, the inaccuracy is below $10\%$. 

\paragraph{Accuracy-cost tradeoff due to sweet spot selection.}

We observe a larger inaccuracy of approximately $30\%$--$40\%$ for the Toric and Surface codes at distances $d=5$ and $d=7$. Although this margin of error is generally acceptable for establishing the order of magnitude, it may be insufficient for applications requiring high-precision estimates. In this section, we investigate the source of this discrepancy and propose a systematic method to reduce it within our framework.

In Section \ref{sec:separating-low-and-high-weights}, we define the sweet spot as:
\begin{equation*}
      y''(\wsweet) \approx \Gamma|y'(\wsweet)|
\end{equation*}
In our previous evaluation, we used the default setting $\Gamma=1$ for all experiments. If we increase $\Gamma$, we shift the sweet spot toward lower weights.
Although this requires a significantly higher testing effort, it allows us to extract more information from low-weight subspaces, resulting in logical error rate estimates with higher accuracy.

As shown in Figure \ref{fig:Tradeoff}, increasing $\Gamma$ consistently increases the accuracy at the expense of higher sampling costs, validating our hypothesis.  We therefore conclude that the low accuracy observed earlier stems from a suboptimal selection of the sweet spot for these two cases. If a more accurate estimate is required and sufficient computational budget is available, one can increase $\Gamma$.

\paragraph{Relative error at different physical error rates.}

\begin{figure*}[ht!]
    \centering
    \begin{minipage}{0.32\textwidth}
        \centering
        \includegraphics[width=\linewidth]{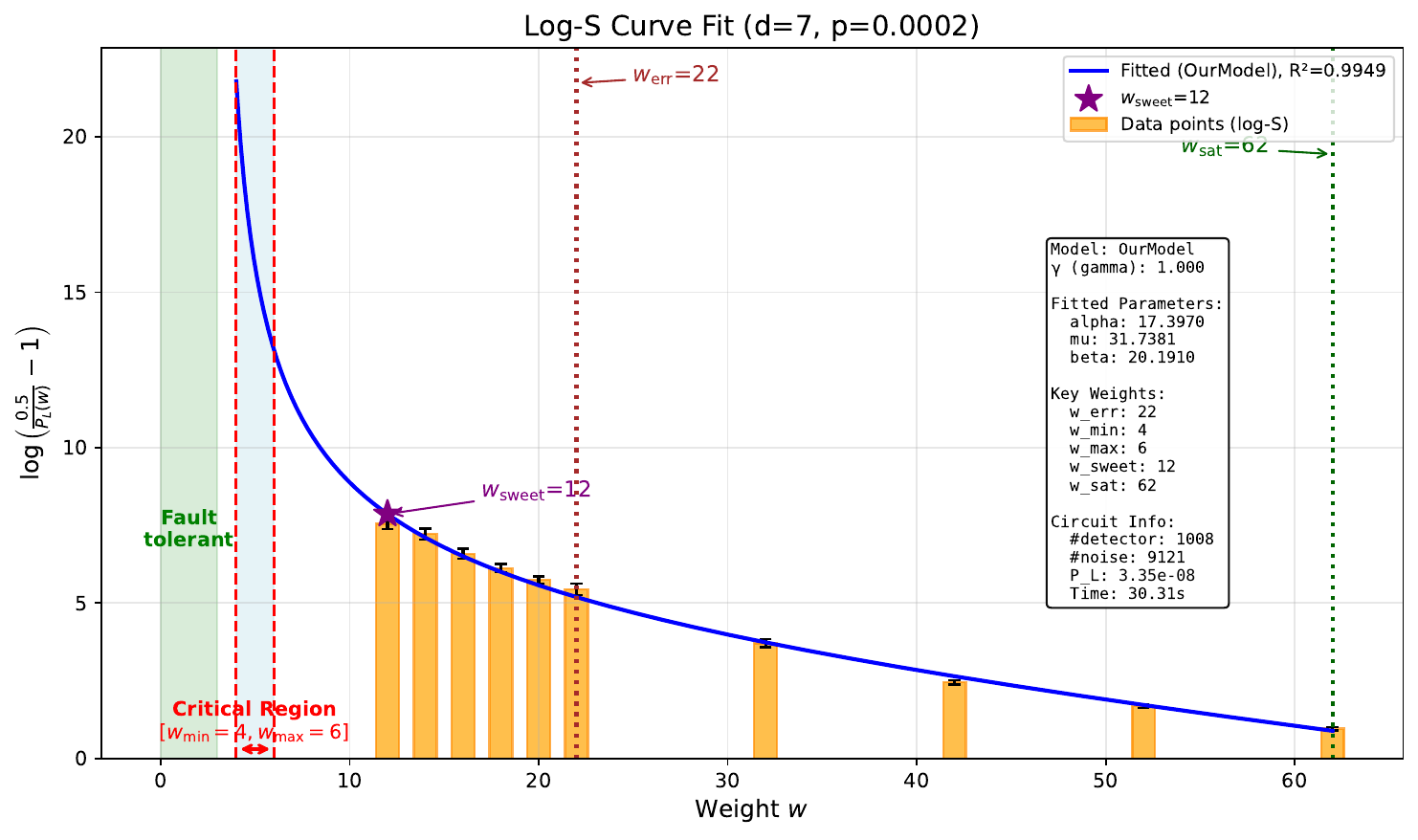}
        \caption*{(a) Surface code d=7, $p=0.0002$}
    \end{minipage}\hfill
    \begin{minipage}{0.32\textwidth}
        \centering
        \includegraphics[width=\linewidth]{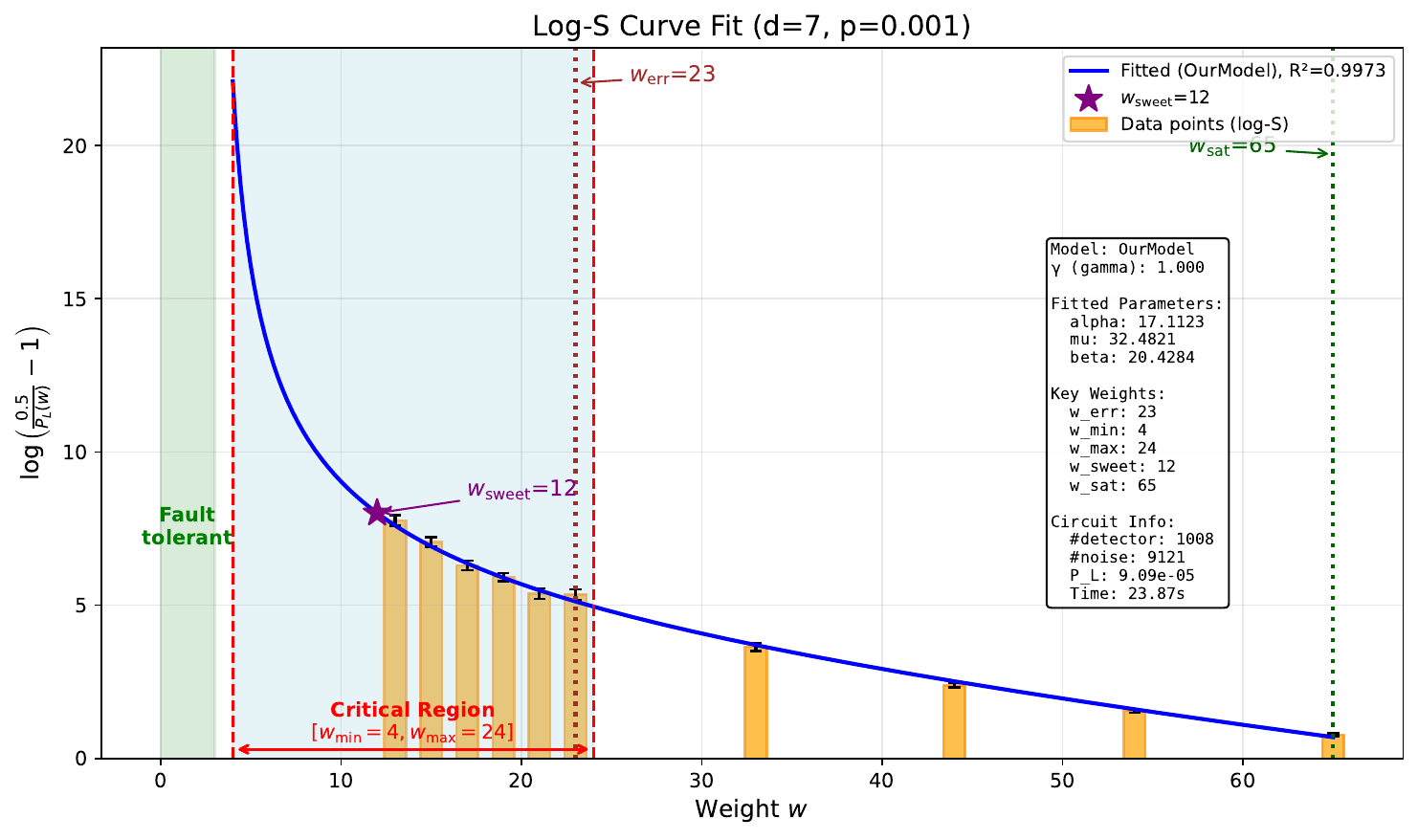}
        \caption*{(b) Surface code d=7, $p=0.001$}
    \end{minipage}\hfill
    \begin{minipage}{0.32\textwidth}
        \centering
        \includegraphics[width=\linewidth]{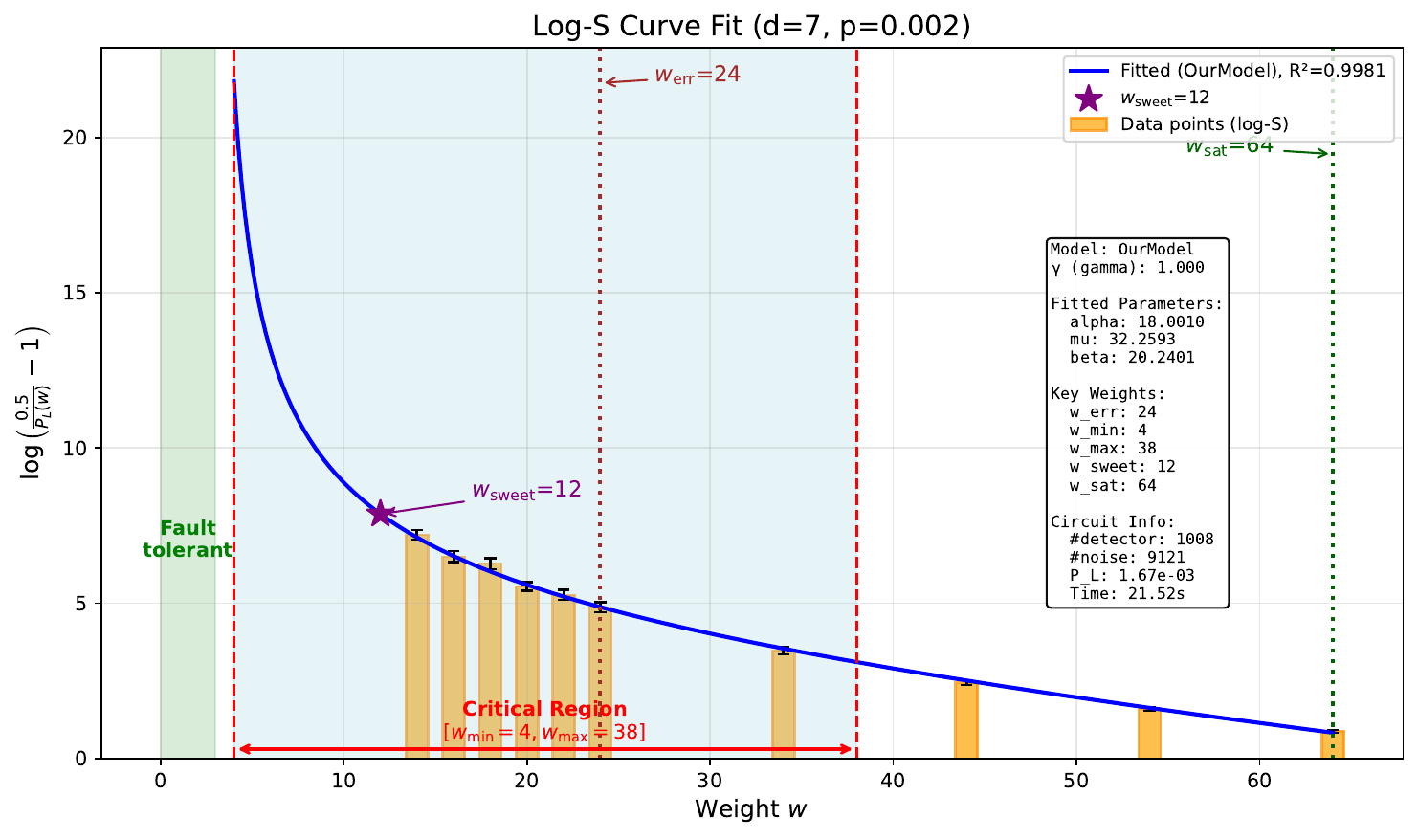}
        \caption*{(c) Surface code d=7, $p=0.002$}
    \end{minipage}
    \caption{The result of testing the same Surface code circuit with distance $7$ under three different physical error rate $0.0002,0.001,0.002$. The blue region highlight the critical weights$[np-5\sigma,np+5\sigma]$ that distribute to logical error rate. Although \ourtool tests enough subspaces towards the sweet spot in all three cases, the accuracy for $p=0.0002$ is much lower than for the other two because our definition of sweet spot is far from the actual critical weights. }
    \label{fig:three-in-a-row}
\end{figure*}

To validate the robustness of our method under different physical error rates, we extend our evaluation over a physical error rate range of $p=0.0001$ to $p=0.01$.

We evaluated both the surface code and the toric codes at distance $d=5$ and $d=7$ using both tools. As illustrated in Figure \ref{fig:errorrate}, we observe an increase in the discrepancy between \ourtool and Monte Carlo sampling using \stim as the physical error rate decreases.

This phenomenon can be explained by the three scenarios shown in Figure \ref{fig:three-in-a-row}. Specifically, for moderate error rate ($p=0.001$ and $p=0.002$), our sweet spot selection accurately fits the curve and extrapolates to the critical subspaces which contribute to the logical error rate. However, when the physical error is much smaller (such as $p=0.0002$), most errors are concentrated at low weights, where the curve extrapolation is less reliable.

\paragraph{Summary.}
Our results show that \ourtool produces estimates that are highly consistent with those of Monte Carlo sampling using \stim. Furthermore, the observed inaccuracy can be explained through a geometric interpretation and can be systematically reduced, given a larger computational budget.

\subsection{RQ3: S-curve Comparison}
\label{sec:DiffScurve}

We compare the effectiveness of IBM's S-curve model, our S-curve model, and three variants of our S-curve model to justify the model design in this research. First, we run our algorithm with IBM's model.

\begin{figure*}[ht!]
    \centering
    \begin{minipage}{0.32\textwidth}
        \centering
        \includegraphics[width=\linewidth]{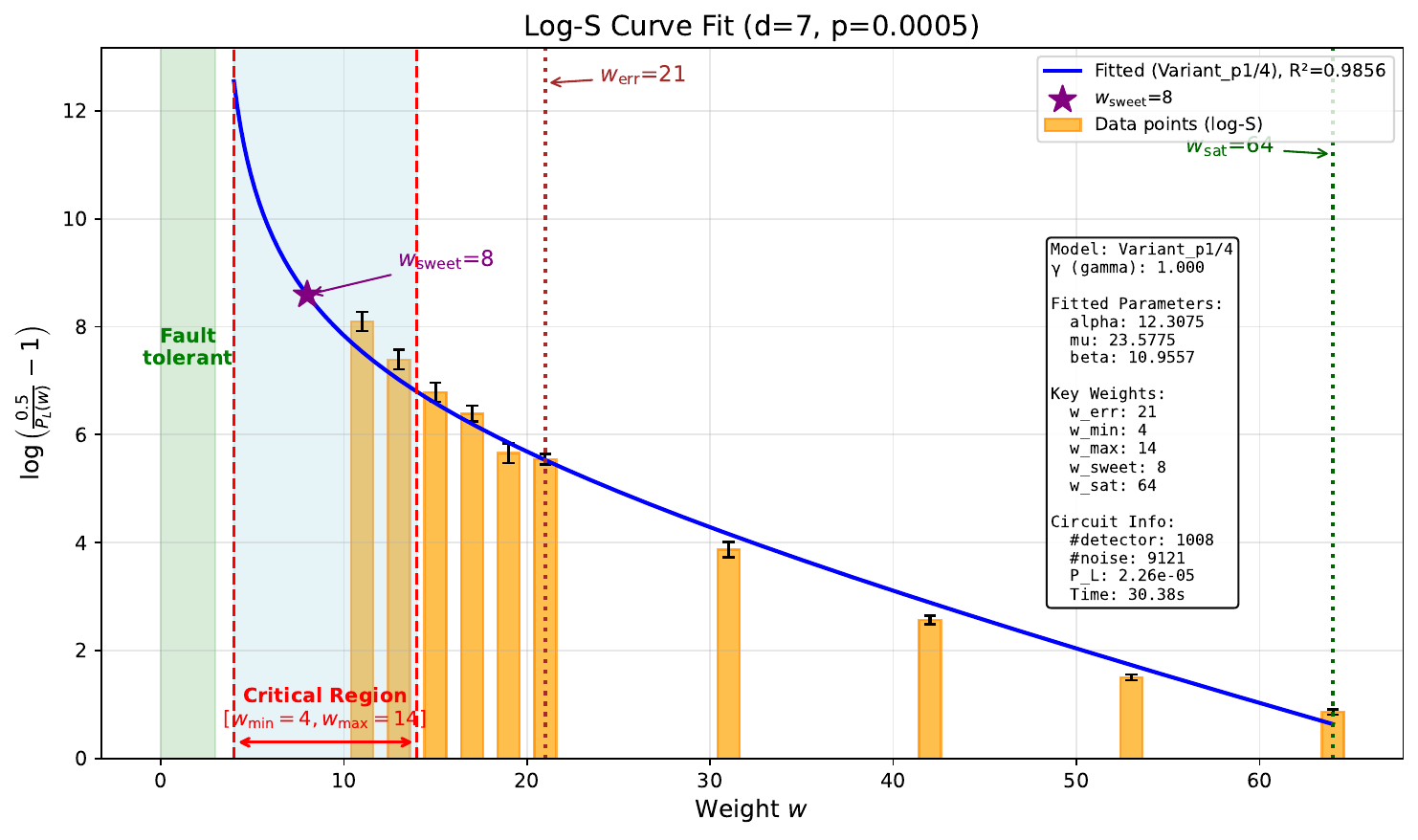}
        \caption*{(a) $s=\frac{1}{4}$}
    \end{minipage}\hfill
    \begin{minipage}{0.32\textwidth}
        \centering
        \includegraphics[width=\linewidth]{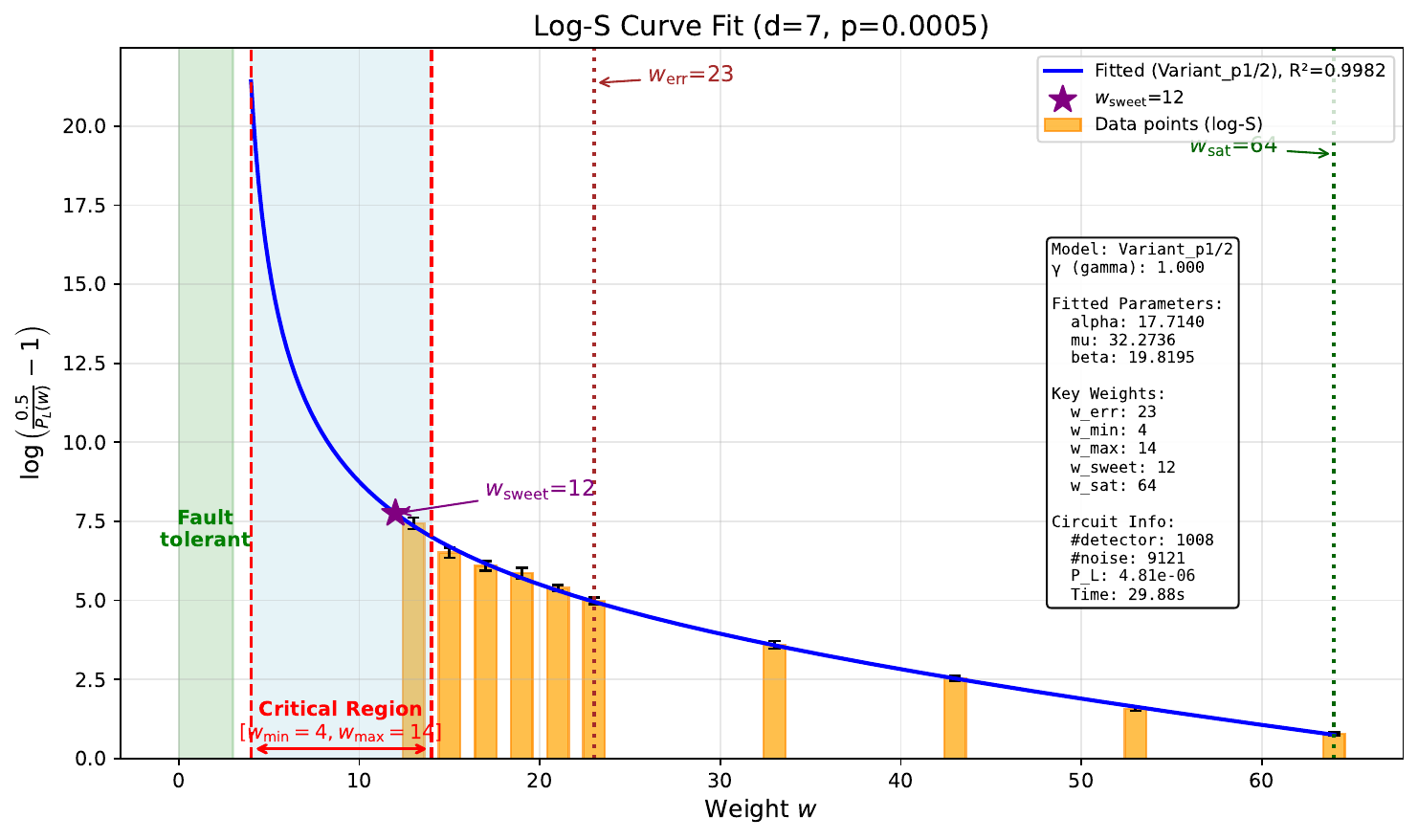}
        \caption*{(b)  $s=\frac{1}{2}$}
    \end{minipage}\hfill
    \begin{minipage}{0.32\textwidth}
        \centering
        \includegraphics[width=\linewidth]{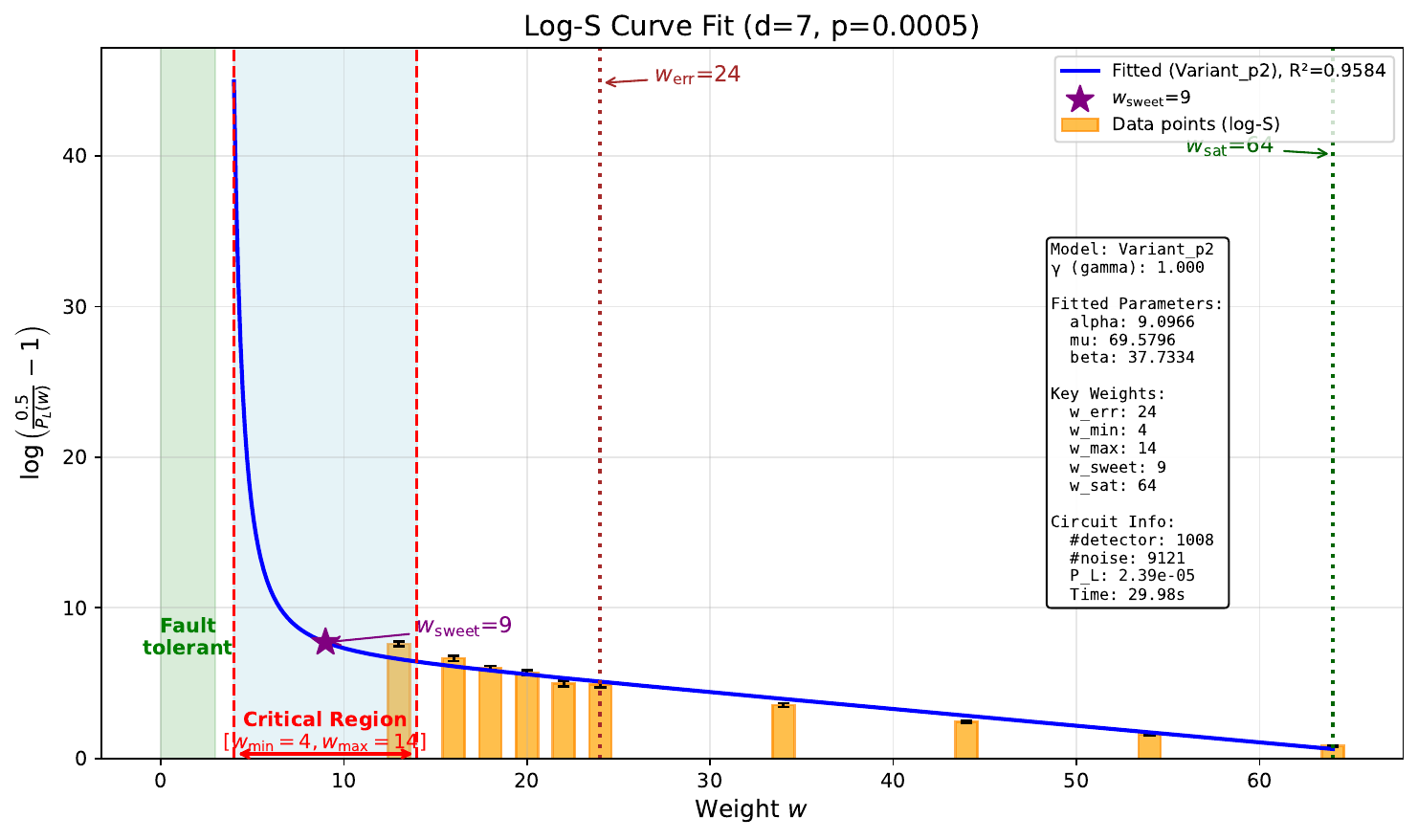}
        \caption*{(c) $s=2$}
    \end{minipage}
    \caption{Illustration of Y-curve fitting on the same Surface code $d=7$ circuit with $p=0.0005$ on different variants of S-curve model. When $s=\frac{1}{2}$, the model extrapolate well for the actual data we observe. }
    \label{fig:three-in-a-row-variantS}
\end{figure*}

\begin{figure}[ht!]
    \centering
    \includegraphics[width=0.45\textwidth]{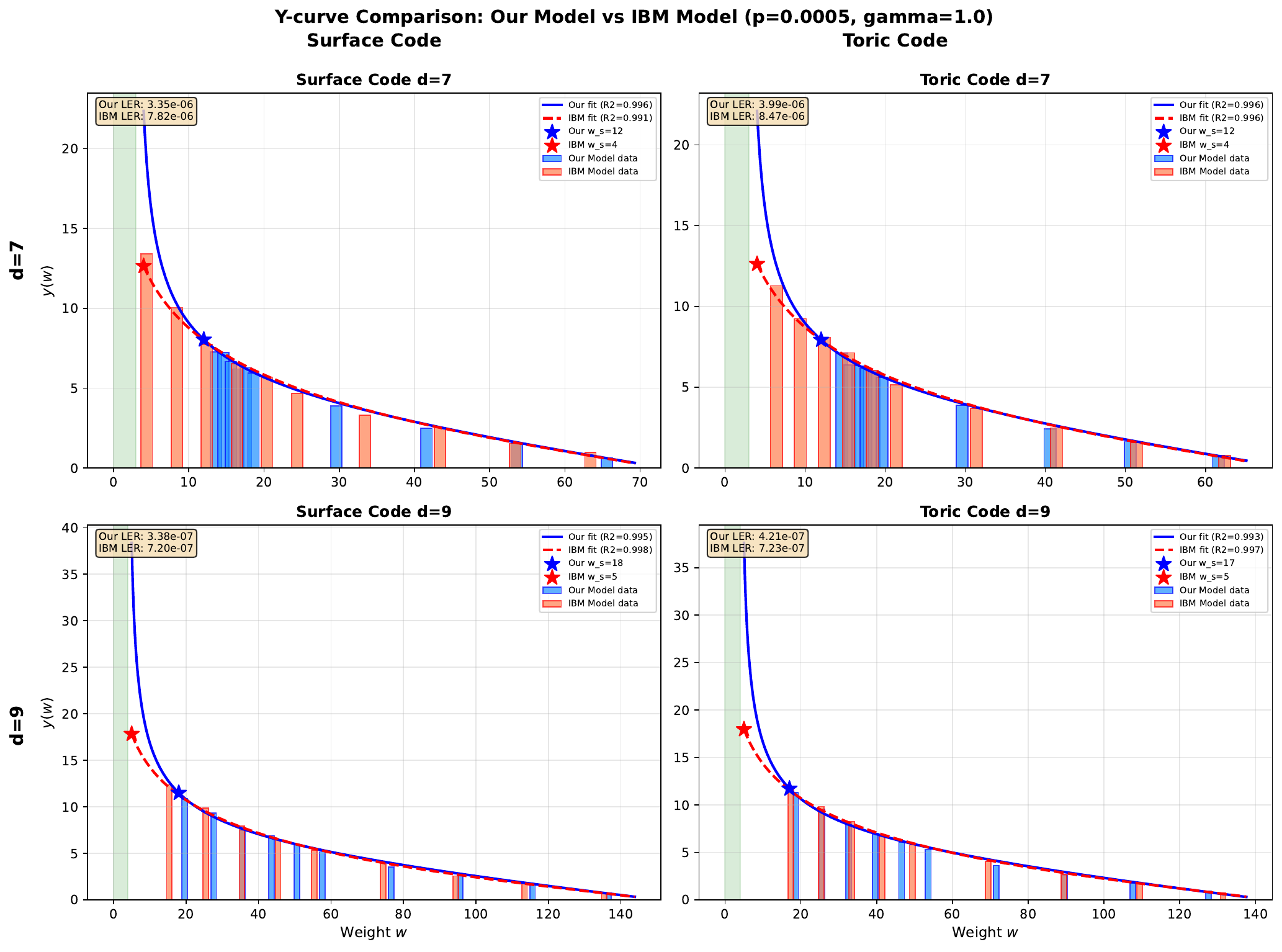}
    \caption{Comparison of fitting and extrapolation between \textcolor{blue}{our S-curve model} and \textcolor{red}{IBM's S-curve model}.}
    \label{fig:compareIBMwithUs}
\end{figure}

\begin{table*}[htbp]
\centering
\small
\resizebox{\textwidth}{!}{%
\begin{tabular}{|c|c|cc|cc|cc|cc|cc|}
\hline
\textbf{Code}
& \textbf{d}
& \multicolumn{2}{c|}{$s=\frac{1}{4}$}
& \multicolumn{2}{c|}{$s=\frac{1}{3}$}
& \multicolumn{2}{c|}{$s=\frac{1}{2}$(Our S-curve model)}
& \multicolumn{2}{c|}{$s=1$}
& \multicolumn{2}{c|}{$s=2$} \\
\cline{3-12}
&
& \textbf{LER} & \textbf{Rel.~error vs MC (\stim)}
& \textbf{LER} & \textbf{Rel.~error vs MC (\stim)}
& \textbf{LER} & \textbf{Rel.~error vs MC (\stim)}
& \textbf{LER} & \textbf{Rel.~error vs MC (\stim)}
& \textbf{LER} & \textbf{Rel.~error vs MC (\stim)} \\
\hline
& 3  & $9.96 \times 10^{-4}$ & $72.7\%$ & $9.91 \times 10^{-4}$ & $71.8\%$ & $5.81 \times 10^{-4}$ & $0.7\%$ & $1.00 \times 10^{-3}$ & $74.1\%$ & $1.06 \times 10^{-3}$ & $83.9\%$ \\
& 5  & $1.34 \times 10^{-4}$ & $108.9\%$ & $1.09 \times 10^{-4}$ & $69.5\%$ & $4.51 \times 10^{-5}$ & $-29.6\%$ & $3.43 \times 10^{-5}$ & $-46.5\%$ & $3.97 \times 10^{-5}$ & $-38.1\%$ \\
Surface
& 7  & $2.38 \times 10^{-5}$ & $300.3\%$ & $1.16 \times 10^{-5}$ & $94.6\%$ & $3.75 \times 10^{-6}$ & $-37.0\%$ & $1.41 \times 10^{-6}$ & $-76.3\%$ & $2.24 \times 10^{-5}$ & $275.2\%$ \\
& 9  & $2.53 \times 10^{-6}$ & $604.7\%$ & $9.34 \times 10^{-7}$ & $160.2\%$ & $3.54 \times 10^{-7}$ & $-1.4\%$ & $6.69 \times 10^{-7}$ & $86.4\%$ & $1.42 \times 10^{-5}$ & $3855.5\%$ \\
\hline
& 3  & $6.38 \times 10^{-4}$ & $-9.5\%$ & $7.52 \times 10^{-4}$ & $6.6\%$ & $7.51 \times 10^{-4}$ & $6.5\%$ & $7.40 \times 10^{-4}$ & $5.0\%$ & $7.18 \times 10^{-4}$ & $1.9\%$ \\
& 5  & $1.95 \times 10^{-4}$ & $134.3\%$ & $1.09 \times 10^{-4}$ & $31.3\%$ & $4.99 \times 10^{-5}$ & $-40.0\%$ & $6.55 \times 10^{-5}$ & $-21.2\%$ & $4.30 \times 10^{-5}$ & $-48.2\%$ \\
Toric
& 7  & $2.39 \times 10^{-5}$ & $250.2\%$ & $1.10 \times 10^{-5}$ & $61.1\%$ & $4.50 \times 10^{-6}$ & $-34.0\%$ & $2.35 \times 10^{-6}$ & $-65.6\%$ & $2.52 \times 10^{-5}$ & $269.2\%$ \\
& 9  & $2.52 \times 10^{-6}$ & $453.4\%$ & $1.39 \times 10^{-6}$ & $205.3\%$ & $4.24 \times 10^{-7}$ & $-7.0\%$ & $8.54 \times 10^{-7}$ & $87.3\%$ & $9.76 \times 10^{-6}$ & $2040.3\%$ \\
\hline
\end{tabular}%
}
\caption{RQ3 (Different S-curve model selection): For each S-curve variant $f_t^s$ in Eq.~\ref{eq:Variants} with $s\in\{\frac{1}{4},\frac{1}{3},\frac{1}{2},1,2\}$, we report the estimated LER and the accuracy relative to Monte Carlo sampling using \stim. When $s=\frac{1}{2}$, Monte Carlo sampling using \stim and \ourtool reach the maximal agreement that we have observed.}
\label{tab:rq3_scurve_variant_p_ler_bias}
\end{table*}

\paragraph{Effectiveness of IBM's model.}

The logical error rates predicted by IBM's model are reported in Table~\ref{tab:ler_comparison_scaler_ibm_bias}. Across all configurations, IBM's model is less accurate than our S-curve model. Figure~\ref{fig:compareIBMwithUs} compares the corresponding Y-curve fits, IBM's model is more conservative when extrapolating from low-weight data, which leads to large overestimates of the logical error rate.

The low accuracy comes from how the two models handle fault-tolerance information.
In IBM's model, the fault-tolerant threshold $t=\frac{d-1}{2}$ is not encoded explicitly,
so the extrapolation toward the fault-tolerant subspaces is intentionally conservative.
In contrast, our S-curve model hard-codes $t=\frac{d-1}{2}$ and constrains the fitted Y-curve
to diverge as $w \to t^{+}$.
These results suggest that explicitly modeling the fault-tolerant threshold substantially increases the accuracy.

\paragraph{Effectiveness of our model.}

Here is a generalization of our S-curve model that takes an additional parameter $s$, which is a positive real number:

\begin{equation}
\label{eq:Variants}
f_t^s[\mu,\alpha,\beta](w)=
\begin{cases}
\displaystyle
\frac{1}{2}\cdot\frac{1}{1+\exp\!\left(-\frac{w-\mu}{\alpha}+\frac{\beta}{(w-t)^s}\right)}
\\[-0.2ex]
\qquad \text{when } t<w,
\\[0.6ex]
0 \qquad \text{when } w\le t.
\end{cases}
\end{equation}

When $s=\frac{1}{2}$, this is the same S-curve as defined in Definition~\ref{def:our-s-curve-model}. We will test $s=\frac{1}{4},\frac{1}{3},1,2$.

Table~\ref{tab:rq3_scurve_variant_p_ler_bias} reports the logical error rate estimated by \ourtool under different S-curve variants, along with the relative inaccuracy with respect to Monte Carlo sampling using \stim. Among the tested setting, $s=\frac{1}{2}$ yields the highest accuracy.

Figure~\ref{fig:compareIBMwithUs} helps to explain this trend. When $s$ decreases to $\frac{1}{4}$, the corresponding Y-curve increases slowly as it approaches the fault-tolerant region, causing \ourtool to overestimate the LER. In contrast, when $s=2$, the Y-curve rises too sharply, making \ourtool underestimate the LER.

\begin{table*}[htbp]
\centering
\small
\resizebox{\textwidth}{!}{%
\begin{tabular}{|c|c|c|c|c|}
\hline
\textbf{Code}
& \textbf{d}
& \textbf{\ourtool LER (IBM Model)}
& \textbf{Rel.~error vs MC (\stim)}
& \textbf{Rel.~error factor 
($\times$)} \\
\hline
& 3  & $1.06 \times 10^{-3}$ & $83.7\%$  & $119.6\times$ \\
& 5  & $9.72 \times 10^{-5}$ & $51.7\%$  & $1.75\times$  \\
Surface
& 7  & $9.61 \times 10^{-6}$ & $61.5\%$  & $1.66\times$  \\
& 9  & $7.52 \times 10^{-7}$ & $109.5\%$ & $78.2\times$  \\
\hline
& 3  & $5.33 \times 10^{-4}$ & $-24.4\%$ & $3.75\times$  \\
& 5  & $1.00 \times 10^{-4}$ & $20.4\%$  & $0.51\times$  \\
Toric
& 7  & $1.17 \times 10^{-5}$ & $71.0\%$  & $2.09\times$  \\
& 9  & $6.77 \times 10^{-7}$ & $48.3\%$  & $6.90\times$  \\
\hline
\end{tabular}
}
\caption{Comparison of LER estimation of Surface and Toric codes using \ourtool with the IBM S-curve model at $p=0.0005$.
The inaccuracy factor is computed as $\bigl|\text{Rel.~error}_{\mathrm{IBM}\ \mathrm{vs}\ \text{MC (Stim)}}\bigr| / \bigl|\text{Rel.~error}_{\mathrm{Our\ model}\ \mathrm{vs}\ \text{MC (Stim)}}\bigr|$, where $\text{Rel.~error}_{\mathrm{Our\ model}\ \mathrm{vs}\ \text{MC (Stim)}}$ is taken from Table~\ref{tab:ler_comparison_all_packed}.}
\label{tab:ler_comparison_scaler_ibm_bias}
\end{table*}

In conclusion, our Y-curve fitting framework is compatible with a broad family of S-curve models. Among the variants we evaluated, our chosen S-curve achieves the highest accuracy.

\section{Threats to Validity}
\label{sec:threats-to-validity}

\paragraph{Do All QEC-Test data Have S-Curves?}
Beverland et al.~\cite{beverland2025failfasttechniquesprobe} proved the S-curve behavior based on a specific set of assumptions. However, there is currently still no rigorous mathematical proof from first principles that all QEC-test data have S-curves. For future QEC circuits, new and better models may be needed in different noise models, decoder algorithms, and compilation schemes.

\paragraph{Elusive Ground Truth.}

We use unconditional Monte Carlo estimates obtained using \stim as ground truth, which works up to distance 10, above which this Monte Carlo baseline does not sample enough logical failures within a two-hour time budget.
For larger distances, the ground truth is elusive, and we gain confidence in the \ourtool output by repeating each experiment and calculating the confidence interval.  %

\section{Related Work}
\label{sec:related-work}

\paragraph{Stim: the State-of-the-Art Tool.}

\stim~\cite{gidney2021stim} is a state-of-the-art and widely used tool to simulate QEC based on randomized fault injection. It is highly optimized at the software and system levels, enabling fast circuit simulation. However, unconditional Monte Carlo sampling becomes sample-inefficient when estimating rare logical failures, even when using an efficient simulator such as \stim. In this paper, we complement unconditional Monte Carlo sampling using \stim by introducing a different approach that scales to large code distances, with a modest time budget.

Why did we not implement our approach inside \stim? \stim is highly optimized for Monte Carlo sampling from the physical noise distribution: it compiles a circuit into a detector error model in which physically distinct faults with the same effect on detectors and observables are merged into a single error mechanism with a combined probability, and its frame simulator draws every mechanism independently, in bit-packed batches, with that probability. Our approach needs the opposite view of the noise. It must enumerate the concrete fault locations of the circuit, draw a fault set of a prescribed Pauli weight uniformly among all fault sets of that size, and propagate exactly that set to the detectors and observables. In \stim, merged or simplified error mechanisms have no fixed weight, and the frame simulator has no notion of a weight stratum, so supporting weight-stratified sampling requires a different system design rather than a new sampling mode. Our QEPG sampler provides that design: a quantum error propagation graph with one node per concrete fault location, from which the detector and observable outcomes of a fault set of any weight follow by propagation, driven by the adaptive weight selection of Algorithm~\ref{alg:ScaLER}. We do use \stim for compiling a noisy circuit into its detector error model, which our decoders consume, and running the Monte Carlo baseline.

\paragraph{Rare-Event Testing of QEC.}

Beverland et al.~\cite{beverland2025failfasttechniquesprobe} proposed three methods to test error correction, namely 1) fitting to a failure-spectrum ansatz, 2) a splitting method with metropolis sampling, and 3) multi-seeded splitting. 
One of their failure-spectrum ansatzes is an S-curve model that appeared earlier in our Definition~\ref{def:ibms-s-curve-model}. 

Our work goes further by providing an end-to-end algorithm for subspace selection and curve fitting under a fixed time budget.
We also give a better S-curve model, an open-source tool, and an experimental comparison with both their failure-spectrum ansatz and with Monte Carlo sampling using \stim.

Carolyn et al.~\cite{mayer2025rareeventsimulationquantum} identify the same challenge and propose addressing the limitations of standard Monte Carlo methods using subset sampling and splitting techniques, similar to the splitting method of Beverland et al.~\cite{beverland2025failfasttechniquesprobe}.

\paragraph{Testing heterogeneous QEC architectures.}
A growing line of work explores \emph{heterogeneous} fault-tolerant architectures that combine multiple code families.
For example, Stein et al.~\cite{Heteral} evaluate a heterogeneous architecture that couples surface-code patches with a QLDPC (``gross'') code.
Rather than performing an end-to-end fault-injection simulation of the full compiled workload, they use an \emph{instruction-level} error model:
they estimate circuit fidelity by \emph{summing} per-operation logical error contributions. %
This compositional estimator necessarily relies on strong assumptions (e.g., independence/locality of logical faults and simplified models for high-weight non-Clifford operators), and it does not capture correlated failure mechanisms arising from code switching, data movement, and scheduling in heterogeneous systems.
At the same time, fully detailed end-to-end simulations are often computationally prohibitive for such architectures due to the complexity of compilation/transpilation and cross-code dynamics.  We leave to future work to apply our approach to heterogeneous QEC architectures.

\paragraph{Formal Methods.}
Wang et al.~\cite{SymbolicQEC} developed a symbolic execution approach to testing QEC circuits, followed by a formal verification method using Coq to test circuit fault tolerance~\cite{huang2025efficient}. However, in contrast to our paper, these approaches have so far been unable to estimate the logical error rate.

\paragraph{Fault Injection in Classical Software Testing.}
Fault injection is widely used in classical software testing.  For example, Feng et al.~\cite{FuzzFault} proposed a coverage-guided fault injection method to evaluate fault recovery in distributed systems. Gao et al.~\cite{CloudInject} developed an efficient fault injection approach for cloud systems, prioritizing mutation combinations based on runtime feedback. Leveugle et al.~\cite{leveugle2009statistical} provided a method to quantify error and statistical confidence in randomized fault injection. Khanfir~\cite{BugReportsGLM} explored generative models and bug reports for fault injection.  In the spirit of those papers and many more for classical computing, \stim is the best-known tool for fault-injection in quantum computing.

\section{Future Work}
\label{sec:future-work}

Our approach uses the error model SID, which is a variant of SD6 \cite{Gidney2021faulttolerant,geher2024error} 
but different from SI1000 \cite{Gidney2021faulttolerant}. 
We leave to future work to enable our approach to take an error model as a parameter.

Our approach uses a hyperparameter $\Gamma$, which we use in the definition of the sweet spot.  We leave to future work to autotune the value of $\Gamma$, which may require a larger time budget than the two hours we use in this paper.

\section{Conclusion}
\label{sec:conclusion}

We have introduced \ourtool, a novel approach and tool to efficiently estimate the logical error rate of QEC circuits. 
Our experiments reveal a consistent S-curve behavior across all benchmark circuits, which we model with a simple, understandable framework. Our S-curve model captures both the circuit's fault tolerance and its performance degradation as the number of errors increase. Crucially, the model's continuity reduces the testing effort required to fit and learn its parameters. \ourtool takes advantage of this insight by integrating model fitting with stratified fault injection, achieving better scalability than Monte Carlo sampling using \stim.

\paragraph{Author contributions.}
John Zhuoyang Ye conceived the approach, wrote the initial ScaLER implementation, conducted the experiments, prepared all figures, and wrote the first draft. Jens Palsberg advised the research and contributed substantially to the writing and revision of the manuscript. After the initial implementation was complete, AI tools assisted with code optimization and bug finding. In preparing the manuscript, the authors used AI for local text edits. OpenAI Codex also assisted with reference checking and preparation of the publication files.

\paragraph{Acknowledgments.}

We thank Quan Do and Keli Huang for helpful comments on a draft of the article.
We thank Craig Gidney for independently testing our distance-17 surface-code estimate with Monte Carlo sampling and for explaining on \href{https://scirate.com/arxiv/2602.04921}{SciRate} why the spread across repeated runs can understate the uncertainty in an extrapolated logical error rate.
We thank the anonymous reviewers at \emph{Quantum} for their thoughtful feedback. Their comments helped us distinguish the sampling cost of estimating rare failures from the speed of the Stim simulator and explain more precisely how ScaLER advances prior work. They also identified an error in the summation range in Eq.~\eqref{eq:surface-logical-error-estimate} and line~5 of Algorithm~\ref{alg:ScaLER}. The corrected sum includes all fault-weight subspaces, improving the accuracy of our logical error rate estimates.
We are supported by the NSF QLCI program through grant number OMA-2016245 and also by NSF grant number 2422170.

\bibliographystyle{quantum-doi}
\bibliography{refs}   %

\onecolumn
\appendix

\end{document}